\documentclass[manuscript]{aastex}

\slugcomment{Submitted to Astronomical Journal}

\shorttitle{LSB galaxy sample from $\alpha$.40-SDSS DR7 survey combination and property statistics}
\shortauthors{Du et al.}

\begin{document}


\title{Low Surface Brightness Galaxies selected from the 40$\%$ sky area of the ALFALFA H{\sc{i}} survey.I.Sample and statistical properties}


\author{
Wei\ Du\altaffilmark{1,3},
Hong\ Wu\altaffilmark{1},
Man I\ Lam\altaffilmark{1,2}, 
Yinan\ Zhu\altaffilmark{1},
Fengjie\ Lei\altaffilmark{1} and
Zhimin\ Zhou\altaffilmark{1}}
\affil{$^1$ Key Laboratory of Optical Astronomy, National Astronomical Observatories, Chinese Academy of Sciences, 20A Datun Road, Chaoyang District, Beijing, 100012, China}
\affil{$^2$ University of Chinese Academy of Sciences, No.19A Yuquan Road, Beijing, 100049, China}
\affil{$^3$ email: {\texttt wdu@nao.cas.cn}}



\begin{abstract}
The population of Low Surface Brightness (LSB) galaxies, which are objects with central surface brightnesses at least one magnitude fainter than the night sky, is crucial for understanding the extremes of galaxy formation and evolution of the universe. As LSB galaxies are mostly rich in gas (H{\sc{i}}), the $\alpha$.40-SDSS DR7 sample is absolutely one of the best survey combinations to select a sample of them in the local Universe. Since the sky backgrounds are systematically overestimated for galaxy images by the SDSS photometric pipeline, particularly for luminous galaxies or galaxies with extended low surface brightness outskirts, in this paper, we above all estimated the sky backgrounds of SDSS images accurately in both g and r bands for each galaxy in the $\alpha$.40-SDSS DR7 sample, using a precise method of sky subtraction. Once subtracting the sky background, we did surface photometry with the Kron elliptical aperture by the SExtractor software and fitted geometric parameters with an exponential profile model by the Galfit software for each galaxy image in the $\alpha$.40-SDSS DR7 sample. Basing on the photometric and geometric results, we further calculated the B-band central surface brightness, $\mu_{0}$(B), for each galaxy and ultimately defined a sample of LSB galaxies consisting of 1129 galaxies with $\mu_{0}$(B) $>$ 22.5 mag arcsec$^{-2}$ and the axis ratio b/a $>$ 0.3 from the 12423 $\alpha$.40-SDSS DR7 galaxies. This H{\sc{i}}-selected sample of LSB galaxies is a relatively unbiased sample of gas-rich and disk-dominated LSB galaxies, which is complete both in H{\sc{i}} observation and the optical magnitude within the limit of SDSS DR7 photometric survey. This LSB galaxy sample spans from 22.5 to 28.3 in $\mu_{0}$(B) with a fraction of 4$\%$ fainter than 25.0 mag arcsec$^{-2}$ in B-band central surface brightness, and distributes from -27.0 to -12.3 mag in the absolute magnitude in B band (M(B)), including 43 faintest galaxies (M(B) $>$ -17.3 mag). This sample is an absolutely blue LSB galaxy sample, of which 98$\%$ are bluer than 0.75 mag in B-V color. As for 21cm H{\sc{i}} properties, our LSB galaxy sample has a median M$_{H{\sc{i}}}$/L$_{B} = $0.87, and a large portion (95$\%$) has high mass of H{\sc{i}} (M(H{\sc{i}}) $>$ 10$^{7.7}$M$\odot$), which supports that galaxies in this LSB galaxy sample are mostly gas-rich. Additionally, we statistically investigated the environment of our LSB galaxies, and found that up to 92$\%$ of the total LSB galaxies have less than 8 neighbouring galaxies, which strongly evidenced that    LSB galaxies prefer to reside in the low-density environment.

\end{abstract}





\section{Introduction}
 Among hundreds of billions of galaxies in the universe, there exists such a class of galaxies called low surface brightness (LSB) galaxy which is at least one magnitude fainter than the night sky in central surface brightness \citep{Impey97,Vollmer13}. In most cases, the central region of a galaxy is the brightest part of the whole galaxy. Thus, a galaxy with its B-band central surface brightness fainter than a certain threshold value is traditionally regarded as a LSB galaxy. However, this threshold value is not unified up to now but it usually varies between 22.0 and 23.0 mag arcsec$^{-2}$ in different papers (e.g. \citet{Impey01,Ceccarelli12}). Besides B band, central surface brightnesses in optical red or even near-infrared band have also been adopted to attempt to distinguish LSB galaxies from high surface brightness (HSB) galaxies, for example, R-band central surface brightness fainter than 20.8 mag arcsec$^{-2}$ \citep{Courteau96,Adami06}, or Ks-band central surface brightness fainter than 18.0 mag arcsec$^{-2}$ \citep{MonnierRagaigne03}.

Limited by the current capacity of observation, LSB galaxies can not be quite easily observed from ground-based optical telescopes due to the faint nature of LSB galaxies themselves and the comparable sky background. Therefore, contributions of LSB galaxies to the universe have been underestimated for a long time, which may further lead a bias to our understanding of the universe.  Due to missing of galaxies with faint surface brightness as a result of the surface brightness selection effect in observation, the faint end of the galaxy luminosity function should be seriously underestimated before \citep{Disney83,McGaugh96}. Therefore, LSB galaxies would play an important role in improving the study of the faint end of galaxy luminosity function \citep{Sprayberry97,Blanton05,Geller12}. Besides to the faint end study of galaxy luminosity function, LSB galaxies are proposed to be significant in many other aspects. For example, they are helpful for us to study the star formation process in the low gas surface density \citep{Noguchi01}, the baryonic matter density and the galaxy formation and evolution \citep{O'Neil00}, and also useful for checking the prediction of the dark matter to the large-scale struction of the Universe \citep{Peebles01}. 

Recently, with the modern techique of observation rapidly improved, more LSB galaxies in the Universe, particularly in the local universe, can be detected. More works have started to focus on studies of the LSB galaxies and found the significance of this type of galaxies on improving our understanding of the Universe. For instance, SDSS is one of the databases well suited for searches for and studies of LSB galaxies. \citet{Kniazev04} identified 129 LSB galaxies with $\mu_{0}$(g) down to 25.0 mag arcsec$^{-2}$ from SDSS Early Data Release (EDR) using their own developed method, among which several unusal LSB galaxies were found, such as one with a structure that looks like two exponential disks and ten new giant LSB galaxy candidates. \citet{Liang07} serendipitously found a new nearby edge-on disk LSB galaxy ($\mu_{0}$(B) $\sim$ 23.68 mag arcsec$^{-2}$) with low metallicity (12 + log(O/H)$\sim$ 7.88) from the SDSS, which was mistakenly classified as a star in the SDSS DR4 database because the ``star'' was actually one of its H{\sc{ii}} regions hosted by galaxies with intrinsically low global luminosities and low surface brightnesses. Then, from the main galaxy sample of SDSS DR4 database,\citet{Zhong08} established a large sample of 12282 nearly face-on LSB galaxies with $\mu_{0}$(B) between 22.0 and 23.6 mag arcsec$^{-2}$. Based on SDSS data of this LSB sample, spectroscopic properties including dust extinction, strong emission-line ratios, oxygen abundances, nitrogen-to-oxygen abundance ratios of LSB galaxies have been statistically studied by \citet{Liang10}, which showed that LSB galaxies with lower surface brightness generally have lower metallicities, dust extinction and stellar mass. Compared with high surface brightness (HSB) galaxies, it was suggested that LSB galaxies have not had dramatically different star formation and chemical enrichment histories from HSB galaxies \citep{Gao10,Liang10}; however, the red and blue LSB galaxies perhaps have different star formation histories:blue LSB galaxies are more likely to experience a sporadic star formation event now while red LSB galaxies are more likely to have formed stars continuously over the past 1-2 Gyr \citep{Gao10,Zhong10}. It is found that LSB galaxies tend to have a lack of companions compared to HSB galaxies at small scales and the isolation is more connected with the survival, formation and evolution of LSB galaxies. When compared to a sample of isolated LSB galaxies, LSB galaxies which are close to neighbors clearly show relatively high star formation signatures and/or have a high population of recently formed stars \citep{Galaz11}.

Recent studies indicated that LSB galaxies contribute 20$\%$ to dynamical mass of galaxies in the universe \citep{Minchin04}, 30$\% \sim$ 60$\%$ to the number density of local galaxies \citep{McGaugh96,Bothun97,O'Neil00,Trachternach06,Haberzettl07}, which suggests that the contribution of LSB galaxies to the universe, especially to the local universe, can not be negligable. Although, Malin-1, the first discovered LSB galaxy, is a giant spiral galaxy, LSB galaxies are in fact dominated by faint galaxies because as high as 95$\%$ of the mass of LSB galaxies are occupied by non-luminous dark matter. They are morphologically extended, disk-like or irregular, and mostly rich in gas, poor in dust, and deficient in metal, and the chemical evolution of LSB galaxies differs from that of galaxies that define the Hubble Sequence \citep{Pustilnik11}. The oxygen abundances of LSB galaxies reach as low as 1/3 $\sim$ 1/5 solor abundance \citep{McGaugh94,Roennback95,Burkholder11}, and so far, the already known galaxies with the lowest metallicities belong to LSB galaxy type.

Proved by practical observations, LSB galaxies have a very small number of H{\sc{ii}} regions, which are also small in size and weak in emission \citep{Schombert92,McGaugh95,Schombert13}. This observational phenomenon indicates that LSB galaxies have very low star formation rates (SFRs) at present \citep{vanderHulst93,vandenHoek00,Galaz11} while they are still abundant in gas material of star formation. Obviously, LSB galaxies are unevolved and the star formation activities of them have been suppressed at a very low level especially in the disk \citep{Mo98,Das09}. Although LSB galaxies have low levels of current star foramtion activities, they contain variaties of stellar populations from very red (B-V $\sim$ 1.2 mag) to very blue (B-V $\sim$ 0.2 mag) in color and nearly covers all the H-R diagram \citep{O'Neil97b,Zackrisson05}. Up to now, the reasons why LSB galaxies have low star formation activities in the condition of still being rich in gas material are still an open question.

LSB galaxies are important populations in the universe, so we need more good samples to better study them. Although the LSB galaxy sample established by \citet{Zhong08} mentioned above has greatly extended the known sample of LSB galaxies at that time, galaxies with B-band absolute magnitude greater than -18.0 mag were cut out from the SDSS DR4 database before LSB galaxy selection. Therefore, the sample of \citet{Zhong08} is mainly composed of intermediate LSB galaxies and inevitably lacks the dwarf LSB ones in the SDSS database. As LSB galaxies are mostly gas-rich, surveys of gas-rich galaxies would provide us an opportunity to select a large number of LSB galaxies, which avoids the surface brightness effects that limit the usefulness of optical surveys for finding LSB galaxies. \citet{Minchin04} assembled a H{\sc{i}}-selected sample of 129 LSB galaxies from a very deep survey for neutral hydrogen (HIDEEP) with the Parkes multibeam system, and estimated the cosmological significance of LSB galaxies which provide more than half of the gas-rich galaxies by number, about 30 per cent of the contribution of gas-rich galaxies to the H{\sc{i}} density of the Universe, about 7 per cent of the contribution to the luminosity density of the Universe, about 9 per cent of the contribution to the baryonic mass density of the Universe, and about 20 per cent of the contribution to the dynamical mass density of the Universe. The ALFALFA survey (Giovanelli 2007) is one of the successful 21cm H{\sc{i}} surveys, and \citet{Trachternach06} performed a blind optical follow-up observations of a part of the region covered by the blind Arecibo H{\sc{i}} Strip Survey (AHISS,\citet{Zwaan97}), and detected optical counterparts of all H{\sc{i}} detections, of which 30 per cent were LSB galaxies at the limiting surface brightness of $\mu_{B,lim}$=25.2$\pm$0.31 mag arcsec$^{-2}$. Fortunately, the data of the 40 per cent sky area of the total ALFALFA survey ($\alpha$.40,\citet{Haynes11}) have been released to the public, and the overlap between $\alpha$.40 and the SDSS ($\alpha$.40-SDSS) would be undoubtfully one of the best survey combinations for us to search a relatively unbiased sample of LSB galaxies. In this paper, we will assemble an order-of-magnitude larger sample of LSB galaxies from the $\alpha$.40-SDSS data, this H{\sc{i}}-selected LSB galaxy sample will be not only complete in H{\sc{i}} observation but also in optical magnitude within the observational limit of SDSS photometric survey, and it would definitely provide a good (relatively unbiased) sample for us to better study the LSB galaxies aligning the SDSS optical, ALFALFA H{\sc{i}} bands and other bands.

As the first paper of our series of papers on LSB work, we will search a H{\sc{i}}-selected sample of LSB galaxies from the overlap between $\alpha$.40 and the SDSS DR7, and make statistical studies about the physical properties of LSB galaxies of our sample, including the optical, radio and environmental properties. Then, for the next few papers we will base on this LSB galaxy sample and study the oxygen abundance, the nitrogen-to-oxygen abundance ratio and perhaps further discuss the origin of the nitrogen element. Combining with the GALEX UV data, the five-color SDSS optical data and the WISE infrared data, and the H$\alpha$ images of the LSB galaxies of our sample, star formation rates of the LSB galaxies will be estimated and we will further study stellar populations to reveal their star formation histories and evolution which is one of the important and still open questions about LSB galaxies.

In this paper, we will give a brief introduction of the ALFALFA survey, $\alpha$.40 catalog, SDSS DR7, and the $\alpha$.40-SDSS DR7 sample in Section 2. The data reduction processes done by ourselves for each galaxy in the $\alpha$.40-SDSS DR7 sample will be described in detail in Section 3, including sky subtraction (Section 3.1), surface photometry by SExtractor (Section 3.2), geometry by Galfit (Section 3.3), and calculation of central surface brightness (Section 3.4). In Section 4.1, we select out a relatively unbiased sample of LSB galaxies from the $\alpha$.40-SDSS DR7 parent sample according to the selection criteria. Then in Section 4, we statistically make studies on optical properties (Section 4.2.1), H{\sc{i}} properties (Section 4.2.2) and the environmental properties (Section 4.2.3) for our LSB galaxy sample. Finally, we summarize our work in Section 5.

\section{DATA}

\subsection{The $\alpha$.40 catalog}
The Arecibo L-band Feed Array (ALFA), a L-band (1.4GHz) receiver in Arecibo, the world's most sensitive radio telescope at Arecibo Observatory in Puerto Rico, enables such an extragalactic H{\sc{i}} survey as ALFALFA (the Arecibo Legacy Fast ALFA survey \citet{Giovanelli05a,Giovanelli07,Haynes07}. Exploiting the large collecting area of the Arecibo antenna and its relatively small beam size ($\sim$3.5$\arcmin$), ALFALFA is a very wide area (7000 deg$^2$ of the sky at high Galactic latitudes) blind extragalacitic H{\sc{i}} survey, which aims to catalog all gas$-$bearing extragalactic objects in the local universe and has conducted a deep and precise census of the local H{\sc{i}} universe over a cosmologically significant volumn. Initiated in February 2005 and completed in 2012, ALFALFA has obtained a H{\sc{i}} line spectral database of more than 30000 extragalactic H{\sc{i}} line sources covering the redshift range between -1600 and 18000 km s$^{-1}$ with a resolution of $\sim$5 km s$^{1}$ \citep{Giovanelli05b}.

Now a catalog of H{\sc{i}} detections covering about 40$\%$ of the full ALFALFA survey sky area, has been publicly released as the $\alpha$.40 catalog \citep{Haynes11}. The sky areas contained in the $\alpha$.40 catalog are regions  07$^{h}$30$^{m} <$ R.A. $<$ 16$^{h}$30$^{m}$, +04$^{\circ} <$ decl. $<$ +16$^{\circ}$, and +24$^{\circ} <$ decl. $<$ +28$^{\circ}$ (the ``spring" region) and 22$^{h}$00$^{m} <$ R.A. $<$ 03$^{h}$00$^{m}$, +14$^{\circ} <$ decl. $<$ +16$^{\circ}$, and +24$^{\circ} <$ decl. $<$ +32$^{\circ}$ (the ``fall" region). The $\alpha$.40 catalog consists of 15855 H{\sc{i}} detections, 15041 of which are certainly extragalactic objects and the left 814 are more likely to be Galactic high velocity cloud (HVC). LSB galaxies are mostly containing large reservoirs of gas (H{\sc{ii}}) \citep {Schombert13}, so the $\alpha$.40 catalog provides us an excellent database to hunt a large number of LSB galaxies in the local Universe.

\subsection{The SDSS DR7}
Mapping one quarter of the entire sky, the Sloan Digital Sky Survey (SDSS; \citet{Gunn98,York00,Lupton01,Stoughton02,Strauss02}) aims to obtain CCD imaging in five broad bands (u,g,r,i,and z) and spectroscopy from 3800 to 9200\AA of millions of galaxies, quasars and stars, using a dedicated wide-field 2.5m telescope \citep{Gunn06} at Apache Point Observatory in New Mexico. Data Release 7 (DR7; \citet{Abazajian09}) is the seventh major data release and provides images, spectra, and scientific catalogs. It is the final data release of SDSS-II, an extension of the original SDSS, which share some footprints with the $\alpha$.40 data set.

\subsection{The Parent sample:$\alpha$.40-SDSS DR7 sample}
By cross-referencing the $\alpha$.40 and SDSS DR7 photometric data sets where the two share footprints, the ALFALFA team provides the cross-identifications of $\alpha$.40 H{\sc{i}} sources with the photometric and spectroscopic catalogs associated with the SDSS DR7 in the $\alpha$.40 catalog. Of the total 15041 extragalactic objects from the $\alpha$.40 catalog, there are 12468 sources having optical counterparts (OCs) in the SDSS DR7 photometric data set \citep{Haynes11}.Here, we have to introduce 2 SDSS photometric catalogs, the PhotoObjAll and the PhotoPrimary catalogs. The PhotoObjAll catalog contains all measured parameters for all photometric objects of SDSS. As an object may be observed two or more times in SDSS imaging due to the overlaps at many levels of the imaging (runs, stripes), the PhotoPrimary catalog has been created to only contain the best observation of an object. Usually, we should take objects in the PhotoPrimary catalog for convincing scientific studies. Among the 12468 ALFALFA sources with SDSS DR7 OCs, there are 12423 sources belonging to the PhotoPrimary catalog. So for a convincing scientific study in this paper, we will only regard these 12423 PhotoPrimary sources as our parent sample.
\begin{figure}
 \includegraphics[scale=0.55]{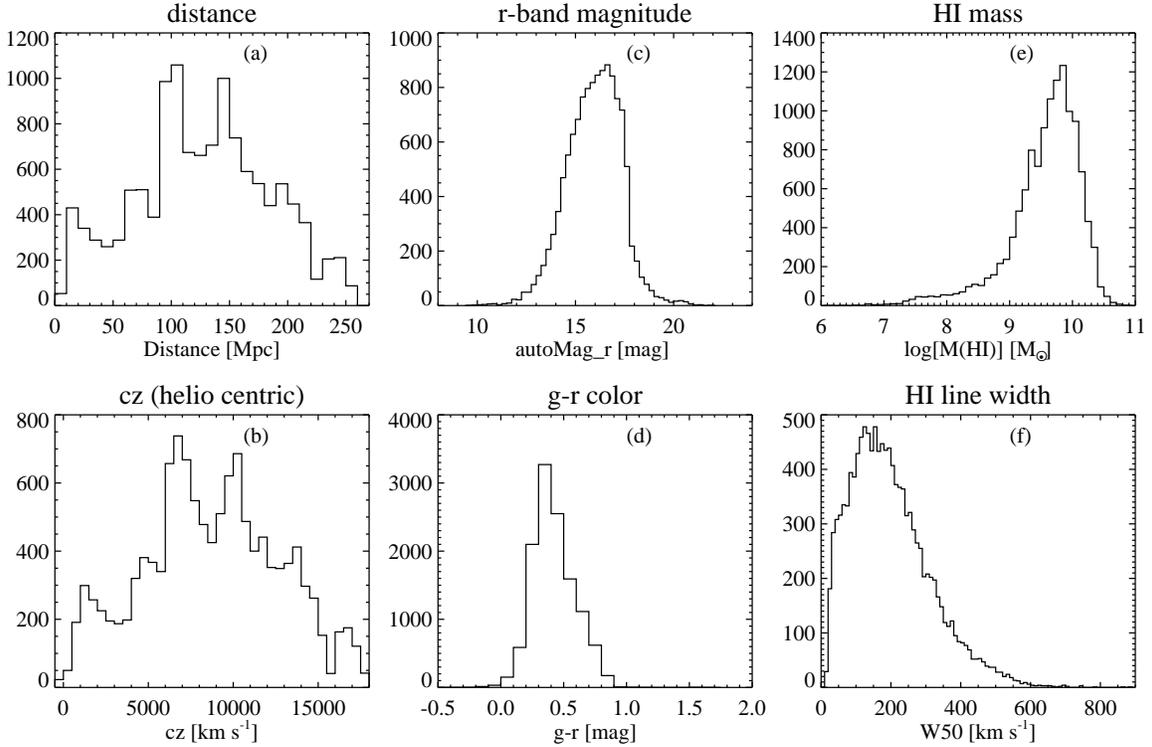}
 \caption[]{Property distributions of the parent sample of 12423 $\alpha$.40-SDSS DR7 sources. The properties are (a)distance with a bin of 10 Mpc, (b)heliocentric radial velocity with a bin of 500 km s$^{-1}$, (c)r-band magnitude with a bin of 0.25 mag, (d)g-r color with a bin of 0.1 mag, (e)H{\sc{i}} mass with a bin of 0.1 dex, and (f)H{\sc{i}} line width with a bin of 10 km s$^{-1}$. Here, magnitudes and colors in (c) and (d) are derived from our own elliptical-aperture photometry described detailedly in Section 3 instead of SDSS photometry \label{fig:fig1}.}
\end{figure}

Here we just give a brief description of this parent sample. As shown in Figure ~\ref{fig:fig1}, the parent sample covers a distance range from 0 to 260 Mpc with 84$\%$ between 50 and 220 Mpc, a H{\sc{i}} mass range from 10$^{6.11}$ to 10$^{10.85}$ M$_{\odot}$ with only 1.6$\%$ as low H{\sc{i}} mass sources (M(H{\sc{i}})$<$ 10$^{7.7}$ M$_{\odot}$; \citet{Huang12}), a heliocentric radial velocity range from -400 to 18000 km s$^{-1}$ with 75.7$\%$ between 5000 and 15000 km s$^{-1}$, a H{\sc{i}} line width range from 9 to 885 km s$^{-1}$ with 99.7$\%$ below 600 km s$^{-1}$, a r-band magnitude range from 9.73 to 21.88 mag with  4.5$\%$ fainter than 18 mag, and a g-r color range from -0.96 to 1.65 mag. Here, the magnitudes of g and r bands and the color of g-r are all derived from the optical surface photometry done by ourselves with elliptical apertures by the SExtractor software \citep{Bertin96}, the process of which will be described detailedly in next section.




\section{Analysis of the optical images of the parent sample}

\subsection{sky subtraction}
SDSS analyzes the raw image data, including bias subtraction, sky and PSF determination, flat-fielding, and finding and measuring the properties of objects by its Photometric Pipeline (PHOTO). For subtraction of the sky background of images, PHOTO performs a strategy of a little simplicity. It at first estimates an initial global sky which is taken from the median value of every pixel in the image, clipped at 2.32$\sigma$. Secondly, PHOTO proceeds to find all the bright objects which are typically those pixels with values more than 51 times of the initial global sky estimation and then mask the bright object pixels. Third, once the bright objects have been masked, PHOTO determines the same clipped median sky value locally within each 256$\times$256 pixel box on a grid with 128 pixels spacing and then bilinearly interpolates this sky value to each pixel to construct a sky background image. Finally, the sky image is subtracted from the image.

This strategy of sky background estimation used in the SDSS PHOTO has its disadvantage, which has a high risk of considering the large extended outskirts of the bright objects as part of the sky background, and thus overestimates the sky background of the bright objects, inevitably causing underestimation of the luminosity and radius of the bright objects \citep{Lauer07,Liu08,Hyde09,He13}. This sky subtraction problem has also been recognized by the SDSS team themselves \citep{Adelman-McCarthy06, Adelman-McCarthy08}, and they quantified the underestimation of the brightnesses of galaxies of large angular extent due to poor sky subtraction to exceed 0.2 mag for galaxies brighter than r=14 mag \citep{Adelman-McCarthy08}. This bias was proved to be even larger in \citet{Lauer07}, which found that for the brightest cluster galaxies (BCGs) in their sample of luminous galaxies, the SDSS luminosities and radius are strongly biased to low values by excessive sky subtraction. For the BCGs with large effective radiuses ($R_{e} > $60 arcsec), the underestimation discrepancies of the SDSS r-band magnitudes resulted from the excessive sky subtraction are even larger than 1.0 mag. \citet{Liu08} performed accurate sky background subtractions and surface photometry for a carefully selected sample of 85 BCGs in the SDSS r band. The comparison of their photometric results with those of the SDSS demonstrated that the SDSS pipeline underestimated the sizes and luminosities of BCGs, and the discrepancies would become larger if the sizes of BCGs were larger.   \citet{Hyde09} compared the SDSS photometric reductions with those of their own code for a subset of their full sample of early-type galaxies and expectedly found that the SDSS underestimated the sizes and brightnesses for large objects because of the sky subtraction problems of the SDSS PHOTO pipeline. \citet{He13} performed accurate sky subtraction and surface photometry for a complete and homogeneous sample of bright early-type galaxies (ETGs) and compared their measurements with those from the SDSS DR7. They found that the SDSS measurements are on average 0.16 mag and maximumly 0.8 mag lower than their own measurements in Petrosian magnitude due to the overestimations of the sky background by the SDSS PHOTO pipeline and smaller in effective radius.  Such underestimations of the luminosities and sizes of the brightest ETGs also led to underestimations of the luminosity density and stellar mass density for bright ETGs. This recognized issue of the sky background subtraction in the SDSS data releases (DRs) from DR1 to DR7 has been significantly improved by reprocessing all SDSS imaging data using a more sophisticated algorithm for sky background subtraction in the SDSS DR8 \citep{Aihara11}. However, comparing with the measurements of \citet{He13}, it still underestimates the luminosities of ETGs by about 0.12 mag on average due to the overestimations of the sky background subtraction in  SDSS DR8.

 Low-luminosity galaxies tend to have low surface brightness than avergae, and the flux of such a galaxy with low surface brightness is significantly reduced because the sky subtraction determination subtracts a substantial fraction of the galaxy light \citet{Strauss02, Blanton05}. For galaxies with extended low surface brightness outskirts which are easily considered as the sky light by the SDSS photometirc pipeline, underestimations for luminosities are still serious, and the bias can even reach as high as 0.5 mag \citep{Lisker07}. So, there are limits of the SDSS sky background subtraction algorithm, not the SDSS data, but deriving a sample of LSB galaxies from photometry alone requires accurate sky background subtraction since an overestimation of background like that in the SDSS PHOTO pipeline would bias the number of the true LSB galaxies towards a higher value.  The photometric pipeline of the SDSS was not optimized for finding low surface brightness objects \citep{Blanton05}. Therefore, if we expect to derive a reliable LSB galaxy sample from the $\alpha$.40-SDSS DR7 parent sample by optical photometry, we have to carefully estimate the sky background images and do the surface photometry ourselves for the SDSS DR7 images of all of the 12423 galaxies in our parent sample.

In preparation, we derive the FITS image files of the corrected frames (fpC-images) with 2048 $\times$ 1489 pixels in both g (4686 \AA) and r (6165 \AA) bands from the SDSS DR7 database for all of the 12423 galaxies in our parent sample. The fpC-images are the images which have been pre-processed by the SDSS photometric pipeline, including bias subtraction, flat-fielding, purge of bright stars and corrections for bad pixels (bad columns, bleed trails, and those corrupted by cosmic rays) by interpolated values. Then, we adopted a more precise method which has developed by \citet{Zheng99, Wu02} to estimate the sky background image for every fpC-image in our parent sample.

Firstly,  we produce a smoothed version of each fpc-image by filtering it with a Gaussian function of FWHM = 8 pixels to make the area of each object a bit more extended in the image (the smoothed fpC-image).

\begin{figure*}
 \centering
 \includegraphics[scale=0.40]{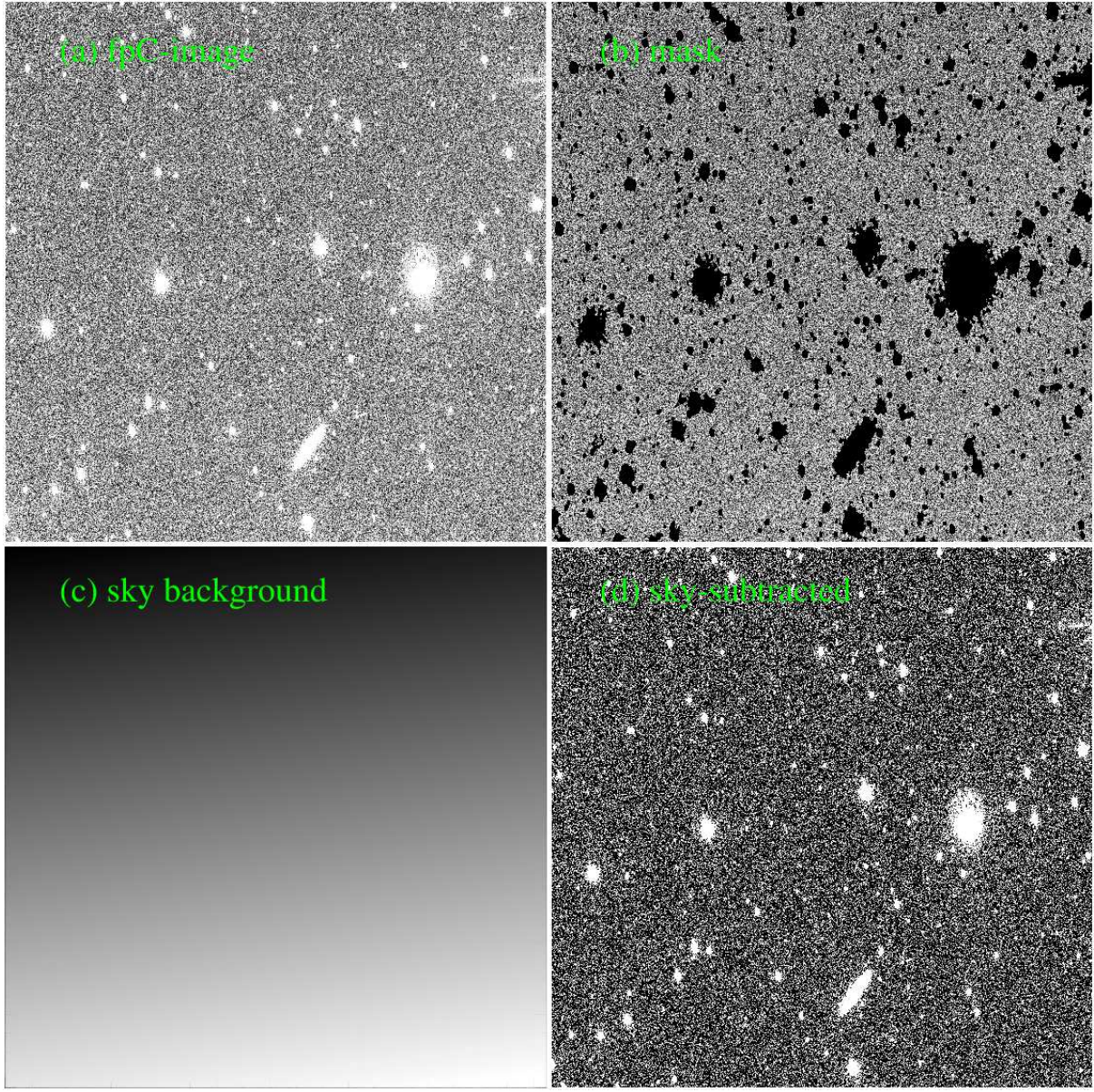}
 \caption[]{Sky subtraction process for a galaxy (AGC 4130) in our parent sample as an example. The objective galaxy AGC 4130, as the edge-on galaxy in this frame locates at the bottom-right region of every picture. The four pictures respectively show (a)the original SDSS DR7 fpC-image in r band for the galaxy, (b)the object-masking file, (c) the sky background map from fitting, and (d)the final sky-subtracted image \label{fig:fig2}.}
\end{figure*}

Secondly, we use the software SExtractor \citep{Bertin96} to automatically detect all objects with the peak flux higher 2.0$\sigma$ above a global sky background value which is simply estimated by the SExtractor itself in the fpC-image. Subsequently, we mask out all the objects detected by SExtractor in the fpC-image to produce an object-masking image. We then carefully check this object-masking image by eye and unfortunately find that the wings of bright stars or the faint stellar haloes of galaxies have not been nicely masked. This would absolutely lead to the overestimation for the sky backgroud and then underestimation for the luminosity of objects. Through trial and error,  we find that if the orginal fpC-image is directly used for SExtractor to detect objects, it may be hard to derive a nice object-masking image which have the wings of the bright objects or the faint stellar haloes of galaxies completely masked. Instead, the smoothed fpC-image generated in last step is an optimal choice for SExtractor to produce a nice object-masking image by nicely detecting objects. It is worth noting that, empirically through trial and error, the Gaussian function with FWHM = 8 seems to be the best choice because it can produce the nicely smoothed fpC-image which can be used by SExtractor to nicely detect not only the bright inner regions of objects but also the just wings of bright objects and the faint stellar haloes of galaxies and then can generate a complete (nice) object-masking file. (e.g. Figure ~\ref{fig:fig2} (b) is the complete mask file for the orginial fpC-image of Figure ~\ref{fig:fig2} (a)).

Thirdly, we subtracted all of the objects from the fpC-image according to the masked areas defined in the complete object-masking file. The object-subtracted image would leave us only but sufficient sky pixels from which we could attempt to precisely determine a reliable sky background.

Lastly, we used the object-subtracted image to model the sky background. We perform a least-square polynomial fit of low order to the sky pixels of each row of the object-subtracted image and replaced the masked pixels of this row by the fitted values. Here, the reason why we restrict the fits to low-order polynomials is that it can avoid introducing spurious fluctuations to the masked regions in the fitted sky by interpolations. This fitting process was performed firstly row by row and then column by column. The individually derived row-fitted and column-fitted sky images were then averaged, and this averaged image was later smoothed with a median filtering box 31 $\times$ 31 pixels in size to eliminate any small artifacts from the modeling process. This smoothed sky images is finally adopted as the sky background (Figure ~\ref{fig:fig2} (c)) and is subtracted from the original fpC-image (Figure ~\ref{fig:fig2} (d) is the sky-subtracted image of the original fpC-image shown as Figure ~\ref{fig:fig2} (a)). Note that the sky backgroud (Figure ~\ref{fig:fig2} (c)) shows a gradient across the frame as expected. Comparing with a straight 2D background fit which has been found to systematically underfit or overfit certain regions of the images \citep{Zheng99}, the method of fitting the sky piecewise in a row-by-row and column-by-column fashion using low-order polynomials can ensure us to predict the sky background underneath the object-masked regions in a mutually orthogonal manner, which not only permits reasonable interpolations of sky under the galaxy region but also fit the inherent lumpiness of the sky at low surface brightness \citep{Zheng99}. Besides, this bidirectional fitting method has already been checked and successfully used in many papers \citep{Zheng99, Wu02, Lin03, Liu05, Duan06, Cao07,Li07, Liu08, Chonis11, Mao14}.

\begin{figure*}
 \centering
 \includegraphics[scale=0.85]{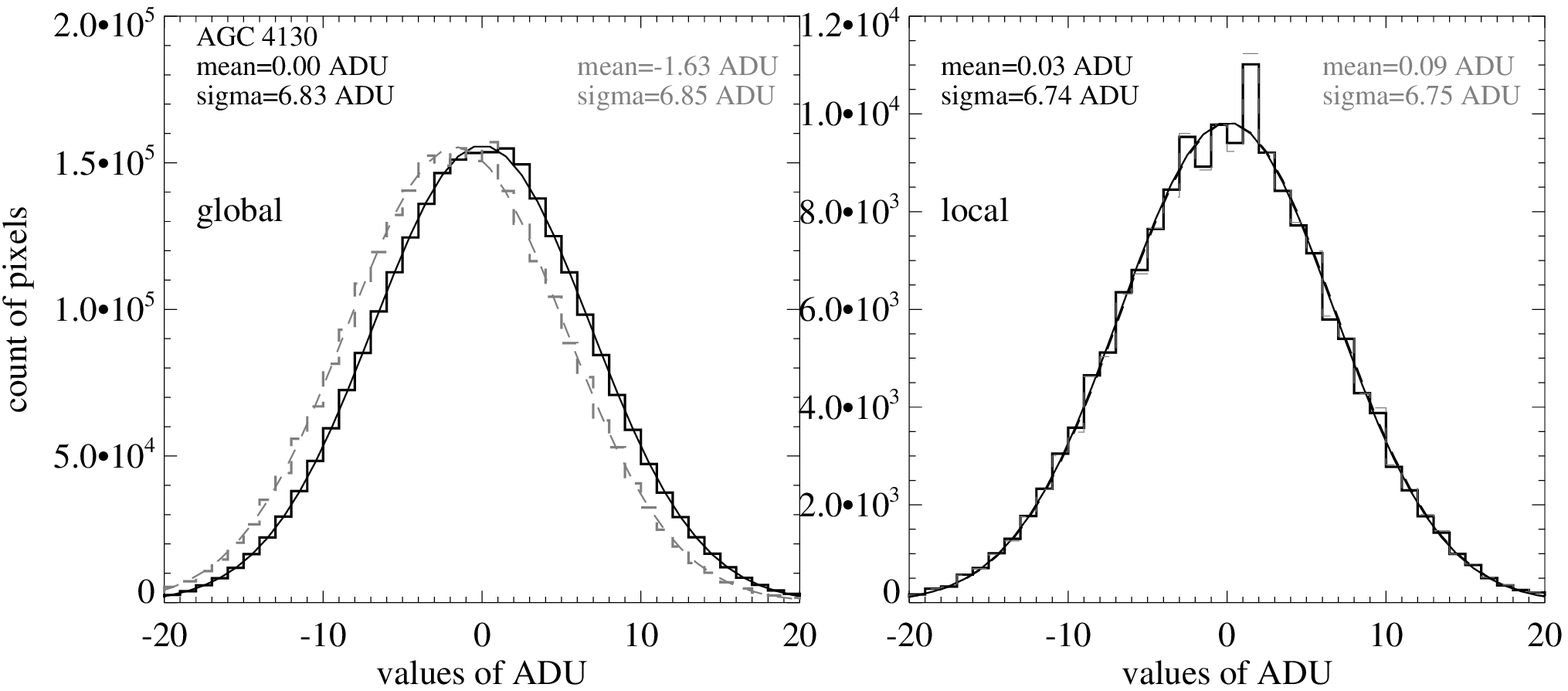}
 \caption[]{Distributions of count for all of the unmasked pixels in the whole frame  (global; the left panel) and in the local vicinity region around the galaxy (local; the right panel). This local vicinity region is defined as the image area between boundaries of two square boxes sized respectively 250$\times$250 pixels and 501$\times$501 pixels from the galaxy center.  The solid black represents distributions for the skysubtracted frame by our piecewise row-by-row and column-by-column fitting method, and, for a further comparison, the dashed grey are for the frame without our accurate sky background subtraction but only with the simple global or local mean value subtraction for all unmasked pixels of the whole frame and of the local region.   \label{fig:fig3}.}
\end{figure*}

Following the specific steps listed above, we have accurately estimated sky background maps and then subtracted the sky backgrounds from the fpC-images for all galaxies in our parent sample in both g and r bands. To check the quality of our sky subtraction, we have made necessary tests.  As an example, for the galaxy presented in Figure ~\ref{fig:fig2}, Figure ~\ref{fig:fig3} shows distributions of counts in the sky-subtracted frame  (the solid black) for all unmasked pixels of the whole frame (global; the left-hand panel) and of the local vicinity around our galaxy (local; the right-hand panel). The around local vicinity is defined by the region between boundaries of two square boxes repectively sized 250$\times$ 250 pixels and 500 $\times$ 500 pixels from the galaxy center. If our sky background model is successful,  the counts are expected to follow a Gaussian distribution with a mean close to zero both globally and locally.  This is indeed the case, as can be seen from the distribution represented in solid black in Figure ~\ref{fig:fig3}. The count distributions for the whole frame and the local region in the SDSS r band are both well fitted by the Gaussian functions with the mean values very close to 0 ADU. The dispersions of the Gaussian distributions are respectively 6.83 ADU for the whole frame and 6.74 ADU for the local region.  Additionally, the distributions represented in dashed grey in Figure ~\ref{fig:fig3} are count distributions in the frame without sky background subtraction by our accurate row-by-row and column-by-column fitting method but only with the simple global or local mean value subtraction for all unmasked pixels of the whole frame and of the local region. These dashed grey distributions are plotted for a further demonstration of the nicety of our sky subtraction. It is clear that the mean values of the Gaussion distributions for the frame with our sky background subtraction (the solid black) are much closer to zero than for the frame with no accurate sky background subtraction but only with the simple global mean value subtraction (the dashed grey) both globally and locally (0 vs. -1.63 ADU for the whole frame and 0.03 vs. 0.09 ADU for the local region). Evidently, this row-by-row and column-by-column method of sky background subtraction can well recover the sky background model, but the most noticable problem of this method, comparing to a straight 2D fitting, might be a little more time-consuming as a result of its relatively complicated fitting algorithm.

\subsection{Photometry by SExtractor}
\begin{figure*}
 \centering
 \includegraphics[scale=0.40]{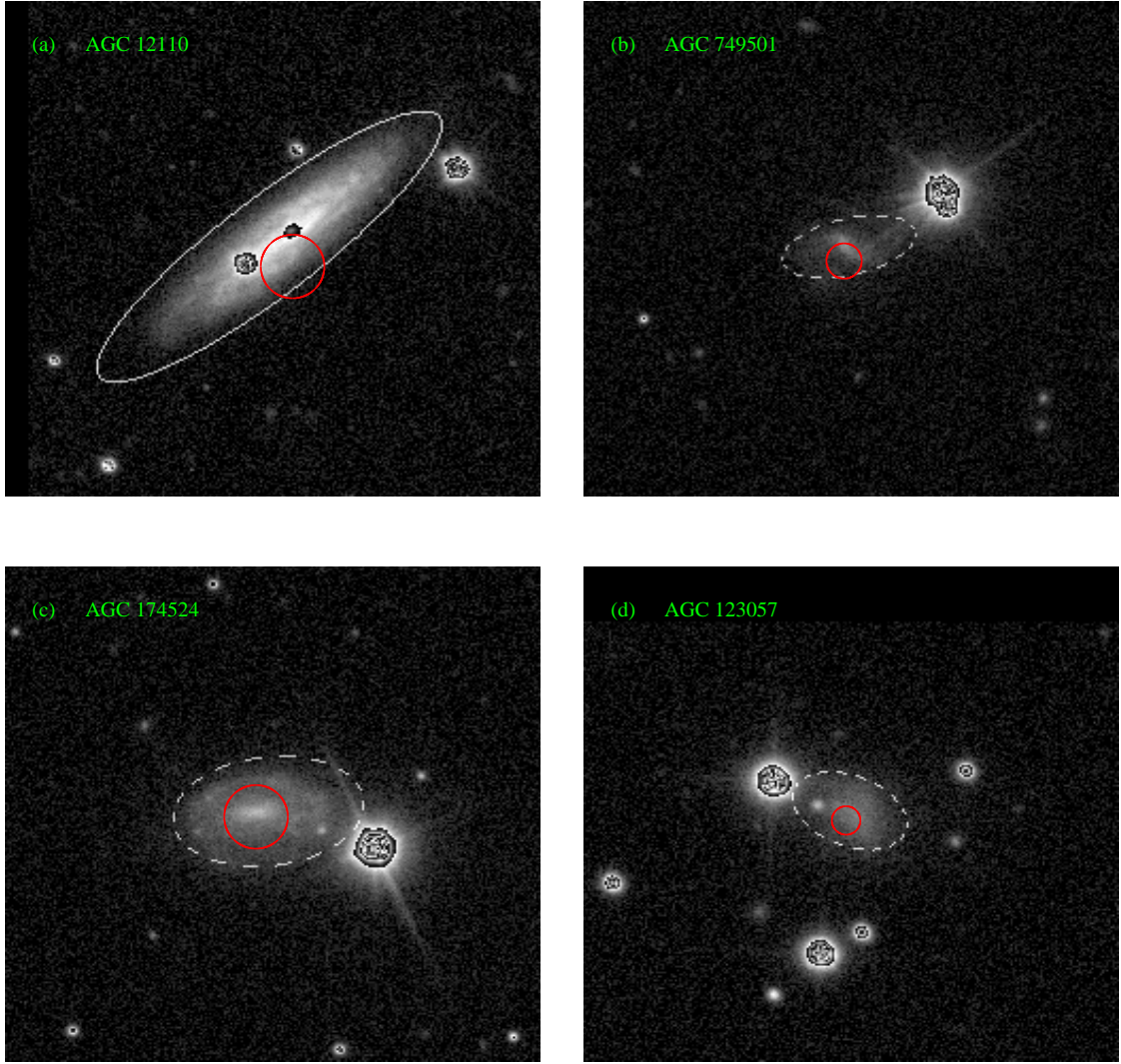}
 \caption[]{Comparisons between AUTO elliptical apertures used by SExtractor and the Petrosian circular aperture used by SDSS in the r band for galaxies from our parent sample.  The white ellipses are the SExtractor elliptical AUTO apertures which are centering at the SExtractor-determined galaxy center. For comparisons, the SDSS cicular Petrosian apertures (red circles) which are centering at the SDSS-determined galaxy centers have been also plotted \label{fig:fig7}.}
\end{figure*}

In the system of the SDSS Petrosian photometry, the Petrosian circular apertures are used \citep{Strauss02}. Although the circular aperture can work well for objects with small angular extent or spherical-like morphology, it is not an optimal choice for galaxies such as most of the galaxies in our parent sample shown in Figure ~\ref{fig:fig7}, which usually have large angular extent, irregular morphology or edge-on shape.  For objects like those shown in Figure ~\ref{fig:fig7}, the circular apertures are too small in size to include all the inherent intrincic light from the objects that we are interested in,  or for objects shown in Figure ~\ref{fig:fig7},  the circular apertures  are so large that they inevitably involve the light from the adjacent objects. As large numbers of LSB galaxies are morphologically similar to late type (Sc and later) spiral galaxies with amorphous or fragmentary and faint spiral patterns, or morphologically similar to irregular galaxies \citep{McGaugh95}. Therefore, the circular apertures are obviously not the best choices for LSB galaxies. In this paper, we will do surface photometry with the elliptical apertures by the SExtractor software \citep{Bertin96} for the sky-subtracted frames of every galaxy in our parent sample. 

The SExtrator package is a source-detection and photometry routine. For photometry, there are six different approaches (isophotal, corrected-isophotal, automatic, best, fixed-aperture, and Petrosian) in the SExtractor software. Thereinto, the automatic aperture magnitudes (AUTO), inspired by Kron's `` first moment" algorithm (see details in \citet{Kron80}), are intended to give the most precise estimate of ``total magnitudes", at least for galaxies. SExtractor's automatic aperture is an flexible and accurate elliptical aperture whose elongation $\epsilon$ and position angle $\theta$ are defined by the second order moments of the object's light distribution. Then, within this aperture, the characteristic radius $r_{1}$ is defined as that weighted by the light distribution function ($r_{1}$= $\frac{\sum rI(r)}{\sum I(r)}$).  \citet{Kron80} and \citet{Infante87} have verified that for stars and galaxy profiles convolved with Gaussian seeing, more than 90$\%$ of the flux is expected to lie within a circular aperture of radius $kr_{1}$ if k=2, almost independently of their magnitudes. This picture remains unchanged if they consider an ellipse with $\epsilon kr_{1}$ and $\frac{kr_{1}}{ \epsilon}$ as the principal axes. By choosing a larger k=2.5, more than 96$\%$ of the flux is captured within the elliptical aperture. So, we keep k=2.5 that is also the default setting of SExtractor during the automatic elliptical aperture photometry \citep{Bertin96}.  More details about the Kron radius and the automatic aperture photometry used in SExtractor can be derived in \citet{Kron80,Infante87,Bertin96}.

Therefore, we would perform the photometry by SExtractor for the sky-subtracted fpC-images in both g and r bands of every galaxy in our parent sample. Firstly, the sky-subtracted fpC-images in r band of all the galaxies in our parent sample were fed into the SExtractor routine \citep{Bertin96} to derive the r-band magnitude and galaxy centeral coordinate of each galaxy. Then, we used the same galaxy center and aperture as those of the r-band image to measure the g-band sky-subtracted fpC-images and derived the g-band magnitude of each galaxy. We show examples of the SExtractor AUTO elliptical aperture at the SExtractor-determined galaxy center (white ellipses in Figure ~\ref{fig:fig7}), comparing with the SDSS circular Petrosian aperture at the SDSS-determined galaxy center (red circles in Figure ~\ref{fig:fig7}) for 4 galaxies in our sample. Obviously, the AUTO photometry appears indeed to be more appropriate for those galaxies because the AUTO ellipses could capture all of the light from the objective sources themselves but exclude the light from other nearby objects. However, the SDSS Petrosian circles adopted in SDSS Petrosian magnitudes seem to have excluded amounts of light from the objective sources themselves or even included light from ajacent sources if the objective galaxy was to be wholely included in a circular aperture.

The photometry for all the galaxies using SExtractor was automatically done in a batch mode, and then we made a visual check on the AUTO aperture photometric results and redid the AUTO photometry by correcting the inappropriate apertures. Finally, the resultant g band magnitude and g-r color are previously shown in Figure ~\ref{fig:fig1} (c) and (d).

\begin{figure*}
 \centering
 \includegraphics[scale=0.45]{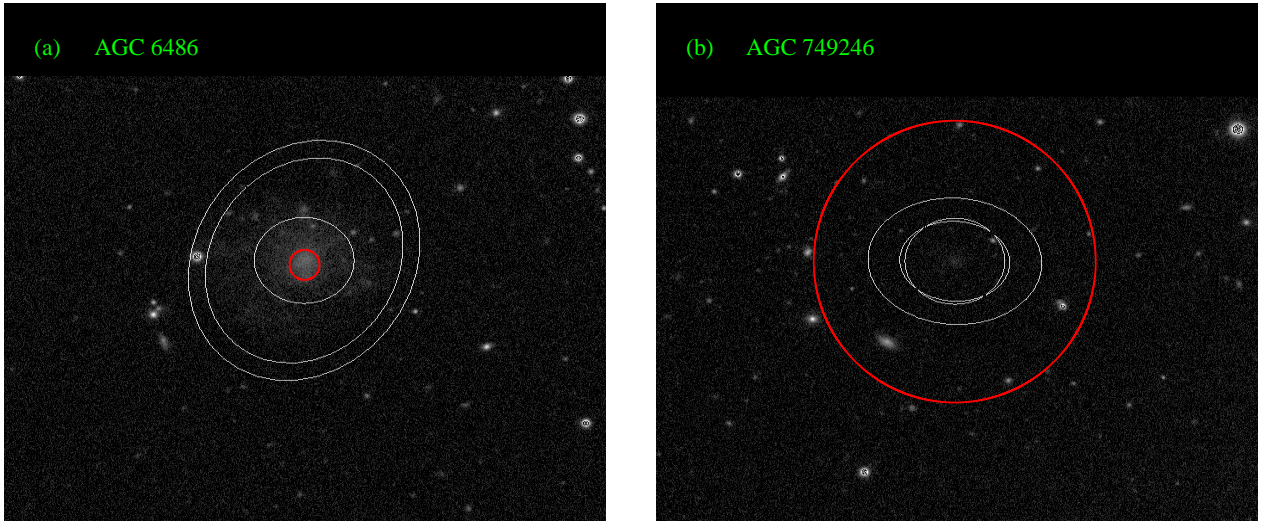}
 \caption[]{Examples of SDSS petrosian aperture (SDSS Petro) and the Kron elliptical apertures (SEx AUTO) from SExtractor for 2 galaxies from our parent sample. The SDSS Petro circular aperture are shown as the red circles, and the SEx AUTO elliptical apertures by the SExtractor routine are shown as the middle white ellipses. Obviously, the SEx AUTO elliptical apertures seem to be much more approriate for the LSB galaxies \label{fig:fig9}.}
\end{figure*}

\subsection{Geometry by Galfit}
Although, besides photometry, some interesting fitted parameters can also be made by SExtractor currently, these fitted results are only the rough guess and estimate of the parameters because the advantage of SExtractor is its photometry instead of geometric fitting by which SExtractor only aims to give a quick look for the interesting fitted parameters. So, we choose to use the Galfit software \citep{Peng02}, which is good at galaxy fitting. By setting a single radial profile function or a combination of a number of functions, e.g. the Sersic, exponential, Nuker models and others which are allowed in most literatures and artifically setting the initial values for a set of input parameters, Galfit starts a nonlinear least squares fitting. During the fitting, minimization of residuals between model and image is done by using the Levenberg-Marquardt downhill-gradient method. The process of minimization iterates until convergence is achieved. Then, the solution in the case of convergence should be regarded as the optimum solution in the parameter space.

\begin{figure*}
 \centering
 \includegraphics[scale=0.5]{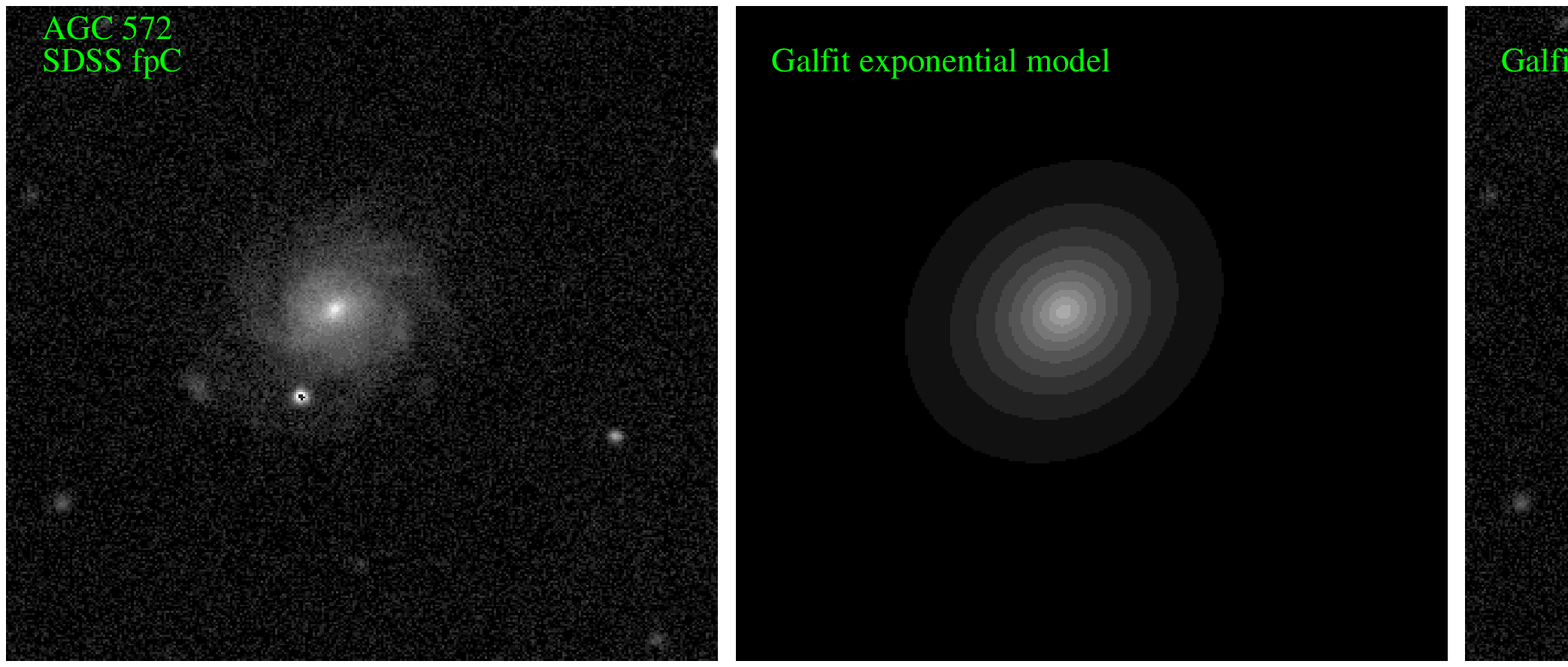}
 \caption[]{Exponential model fit to the galaxy AGC 572 of our parent sample by the Galfit routine. The left, middle, and right panels are respectively the original galaxy image (SDSS g-band), the exponential profile model, and the residual image.\label{fig:fig10}.}
\end{figure*}

Therefore, after surface photometry by the SExtractor software, all of the sky-subtracted images in g and r bands of the 12423 galaxies of our parent sample were fed into the galaxy-fitting procedure Galfit \citep{Peng02} in a batch mode. As majorities of LSB galaxies lack strong bulges \citep{deBlok95,Beijersbergen99}, a decomposition into bulge and disc is not essential for LSB galaxies, especially for the more disk-dominated LSB galaxies which are preferred in our future science goal.  So, we fit each galaxy in our parent sample only with an exponential profile function, and set the parametric results from the SExtractor to be the initial values for the set of input parameters (galaxy magnitude, disk scale length, b/a ratio and position angle) for Galfit, which would finally determine the best-fit values for the interesting parameters including the disk scale length in pixel($\alpha$), axis ratio ($q$), and inclination angle ($i$) for each galaxy of our parent sample and generate a triplet of thumbnails including the original galaxy, the exponential model fit and the residual image (see Figure ~\ref{fig:fig10}).

Although the best-fit results are derived in the case of convergence and $\chi^2$ minimization,  problems with local minima and numerical degeneracies between some parameters may be present as the parameter space is large \citep{Peng02}.  To reinforce our confidence in the parameter optima, we use Monte Carlo simulations to search the parameter space for each galaxy of our parent sample. We make 200 Monte Carlo simulations for each galaxy of our parent sample by the following steps. \\
1. Set the exponential function as the galaxy radial profile function for Galfit.\\
2. Randomly select a set of parameters (magnitude, disk scale length, b/a ratio, position angle), drawing from distributions of these parameter values centered on the best-fit values previously determined by Galfit fitting.\\
3. Using these randomly-selected parameters as initial guesses for the set of input parameters, minimize the $\chi^2$ until the convergence is reached by Galfit.\\
4. Repeat steps 1-3 for 200 times to see whether they return the same optimized solution as that derived by using the SExtractor results to be the initial guesses for Galfit or to other equally plausible ones.\\

\begin{figure*}
 \centering
 \includegraphics[scale=0.8]{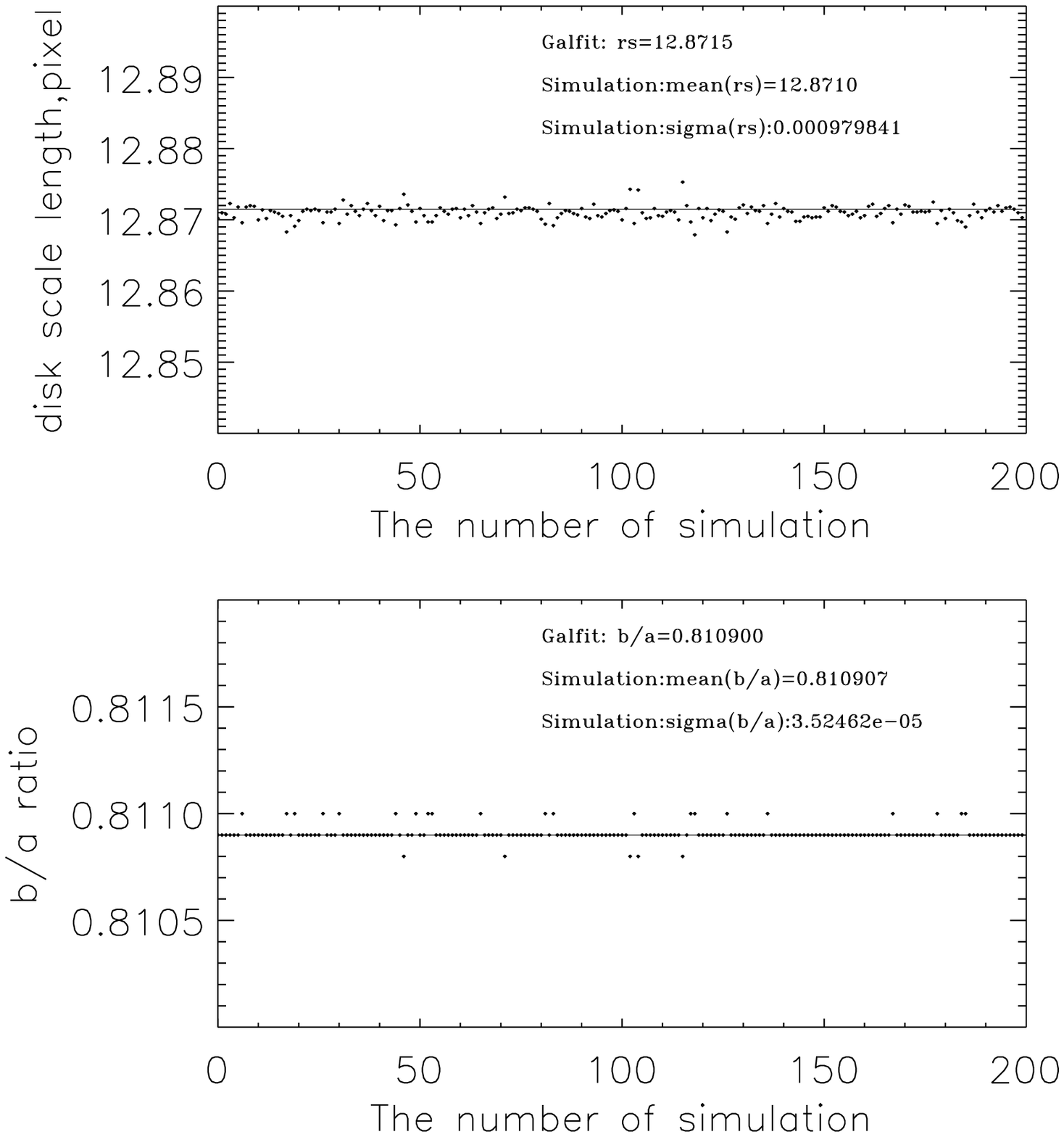}
 \caption[]{Comparison between the ``best-fit solution" and the ``simulated solutions" from the Galfit routine for the galaxy AGC 572. From top to bottom is respectively the comparison for the useful parameters of disk scale length (rs) and axis ratio (b/a). The ``best-fit solution" is represented by the solid line in each panel and the dots are representing the ``simulated solutions" from 200 Monte-Carlo simulations. In the bottom panel,  b/a values seem to be only three discrete values of 0.8108, 0.8109, and 0.8110. This is because that the Galfit software has rounded it final output results to only four decimal places, any difference after the four decimal places can not be apparent. \label{fig:fig14}}
\end{figure*}

For convenience, we use the term of ``best-fit solution" to denote the best-fit solution from Galfit by using the SExtractor results as the initial guesses, and the term of ``simulated solutions" to denote the solution from Monte-Carlo simulations by Galfit which use randomly-selected parameters as the initial guesses. We compare the ``best-fit solution" with the ``simulated solution" for galaxies of our parent sample and find good agreement between the two solutions for every galaxy, which strongly verifies the validity of our best-fit solutions for the useful galaxy parameters. For an example, in Figure ~\ref{fig:fig14}, we take the same galaxy (AGC 572) as that shown in Figure ~\ref{fig:fig10} which shows the galaxy-fitting process of Galfit to present the comparison between the ``best-fit" and the ``simulated" solutions. In Figure ~\ref{fig:fig14}, comparisons between the ``best-fit solution" (the solid line) and the ``simulated solutions" from 200 Monte-Carlo simulations (the dots) are shown for the two useful parameters in this paper of the disk scale length and axis ratio respectively from top to bottom. For each panel from this figure,  the ``simulated solutions" from 200 Monte-Carlo simulations strongly converge at their mean value (denoted as ``Simulation mean and sigma" in Figure ~\ref{fig:fig14}) and the mean value of the ``simulated solutions" is well consistent with the ``best-fit" solution from Galfit (denoted as ``Galfit" in Figure ~\ref{fig:fig14}). \textbf{It is worth noting that in the bottom panel of Figure 7, it looks like that the axis ratio b/a values from simulations are 0.8108, 0.8109, and 0.8110 only,  with nothing in between. Acturally, it is because that the Galfit software has rounded its final output results to only four decimal places, so any difference between the b/a values after four decimal places can not be apparent to us. This further strengthens the consistency of the b/a values from 200 Monte-Carlo simulations as the differences are at least after three decimal places.}

\subsection{Central Surface Brightness}
The typically used photometric parameter to separate the high and low surface brightness regime of galaxies is the central surface brightness of the disc in the B band, $\mu_{0}$(B) \citep{Freeman70,Impey96,O'Neil97a,Zhong08}. So in this subsection, we will calculate $\mu_{0}$(B) for each galaxy in our parent sample. Deduced from the exponential function of the radial profile of the disk component of a galaxy, the central disk surface brightness can be expressed as eq.(1a) where  $\mu_{0}$ refers to the central disk surface brightness in mag arcsec$^{-2}$, $m$ refers to the apparent magnitude in mag, $\alpha$ represents the disk scale length in pixel and $A$ is the area of one pixel. For the SDSS image, $A$ equals to 0.396$\times$0.396 in arcsec$^{2}$/pixel. Following \citet{O'Neil97a,Trachternach06,Zhong08}, the central surface brightness should furthermore be corrected by inclination and cosmological dimming effects, so it can be finally expressed as eq.(1b), where $q$ and $z$ refer to the axis ratio and redshift, respectively.


\begin{mathletters}
\begin{eqnarray}
 \mu_{0}=m+2.5\log_{10}(2\pi \alpha^2A)\\
\mu_{0}=m+2.5\log_{10}(2\pi \alpha^2Aq)-10\log_{10}(1+z)\\
 \mu_{0}(B)= \mu_{0}(g)+0.47(\mu_{0}(g)- \mu_{0}(r))+0.17
\end{eqnarray}
\end{mathletters}


Using the magnitude measured with the AUTO aperture by SExtractor and subsequently Galactic-extinction corrected by us as $m$, the disk scale length in $pixel$ ($\alpha$), the axis ratio ($q$) from the exponential profile fittings by Galfit, the redshift ($z$) deduced from the H{\sc{i}} velocity in the $\alpha$.40 parametric catalog, and the pixel area of $A$= 0.396$\times$0.396 in $arcsec^{2}/pixel$ for the SDSS image, we firstly calculated the central disk surface brightness in both g and r bands, $\mu_{0}$(g) and $\mu_{0}$(r), in units of mag arcsec$^{-2}$ according to Eq.(1b). Then, the B-band central disk surface brightness was calculated from $\mu_{0}$(g) and $\mu_{0}$(r) by using (eq.1c), which stems from the filter transformation relation between g, r and B band (Eq.(2a)) derived from \citet{Smith02}.

\section{Low Surface Brightness Galaxies}
\subsection{LSB galaxy sample}
LSB galaxies are commonly defined as a population of diffuse galaxies whose central surface brightness in B band, $\mu_{0}$(B),
falls below a specific threshold value, which is 21.65$\pm$0.30 in \citet{Freeman70}, 22.0 in \citet{O'Neil97a}, and 23.0 in \citet{Impey97}. In general, the most common threshold values found in the literature are between 21.5 and 23.0 mag arcsec$^{2}$. In this paper, we adopted a threshold of $\mu_{0}$(B) $\geq$ 22.5 mag arcsec$^{-2}$ for the B-band central surface brightness and put additional constraints on the axis ratio (b/a $\geq$ 0.3) to select a sample of LSB galaxies with no edge-on galaxies. With these two constraints, we finally constructed a sample of LSB galaxies consisting of 1129 non-edge-on LSB galaxies from the $\alpha$.40-SDSS DR7 survey. This sample is a relatively unbiased sample of the disc-dominated LSB galaxies and it is complete both in H{\sc{i}} observation and in optical magnitude within the magnitude limit of SDSS photometric observations.
\subsection{Statistical properties of LSB galaxy sample}
 \subsubsection{Optical properties}
Distributions of the B-band central surface brightness ($\mu_{0}$(B)), absolute magnitude (M(B)), the B-V color (B-V), and the scatter of $\mu_{0}$(B) against B-band apparent magnitude (B) for all galaxies in our LSB galaxy sample are shown in Figure ~\ref{fig:fig11} (a)$\sim$(d).

The B-band central surface brightness of all these LSB galaxies (Figure ~\ref{fig:fig11} (a)) are distributing from 22.5 to 28.3 $mag arcsec^{-2}$, with 53$\%$ (597/1129) betweeen 22.5 and 23.0 $mag arcsec^{-2}$, 34$\%$ (388/1129) between 23.0 and 24.0 $mag arcsec^{-2}$, 9$\%$ (103/1129) between 24.0 and 25.0 $mag arcsec^{-2}$, and the left 4$\%$ (41/1129) with the B-band central surface brightness fainter than 25.0 $mag arcsec^{-2}$.

The coverage of the B-band absolute magnitude (Figure ~\ref{fig:fig11} (b)) of all the galaxies in the LSB galaxy sample are from -27.0 to -12.3 $mag$. There are 43 galaxies in our LSB galaxy sample being instrinsically very faint ($M$(B)$>$-17.3 $mag$; \citet{Poggianti01}). We present the distribution of the B-V color of the LSB galaxy sample in Figure ~\ref{fig:fig11}(c), which reveals that our sample covers from -0.22 to 1.88 mag in B-V color and it has 98.4$\%$ galaxies bluer than B-V=0.75 $mag$. This strongly probes that this sample is a bluish sample of LSB galaxies.
Additionally, we depicted the distribution of B-band central surface brightness againt apparent magnitude in Figure ~\ref{fig:fig11} (d), from which we can see a trend that fainter galaxies tend to have fainter central surface brightness.

\subsubsection{21cm H{\sc{i}} properties}

\begin{figure*}
 \centering
 \includegraphics[scale=0.45]{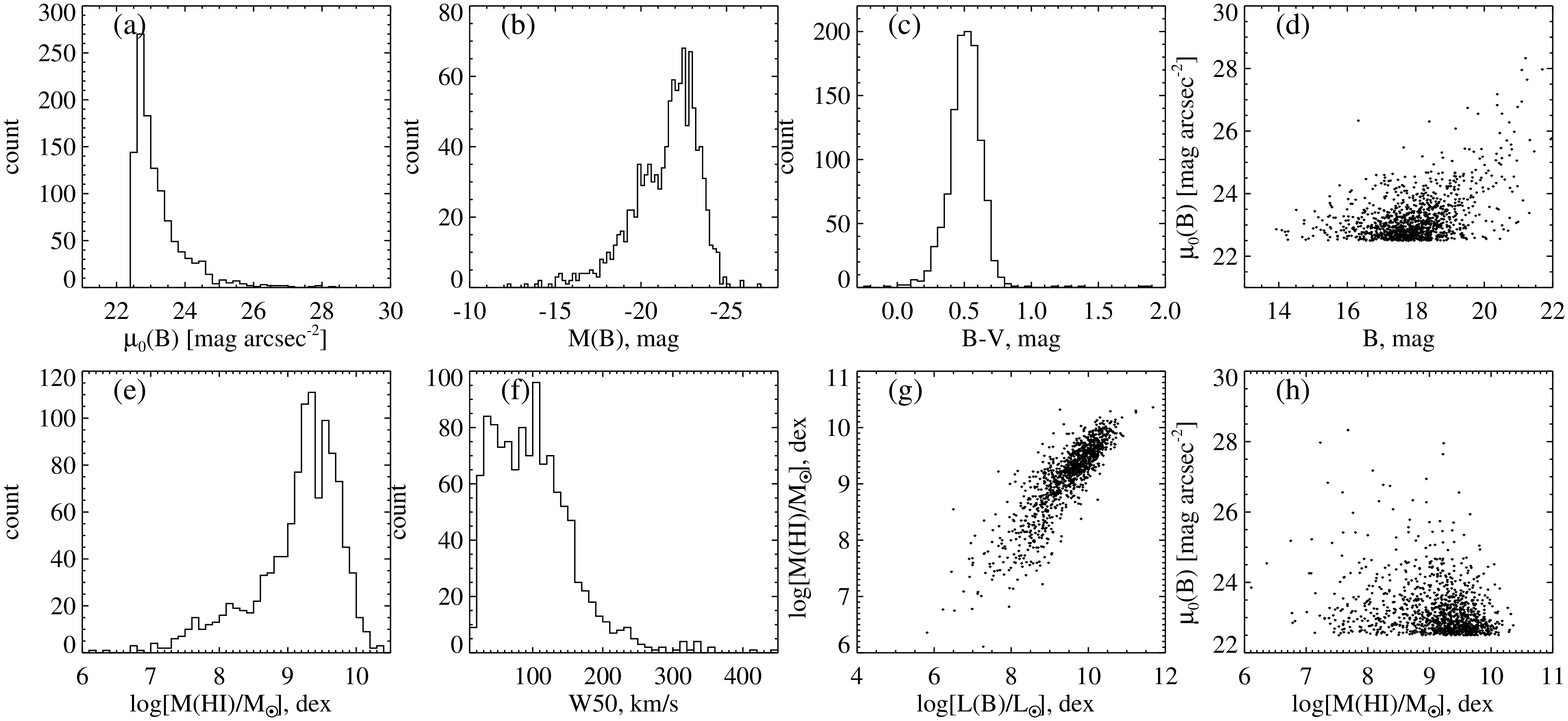}
 \caption[]{Optical and H{\sc{i}} properties of the full LSB galaxy sample. Panels respectively represent distributions of (a) B-band central surface brightness with a bin of 0.2 $mag arcsec^{-2}$, (b) B-band absolute magnitude with a bin of 0.2 $mag$, (c) B-V color with a bin of 0.05 $mag$, (d) central surface brightness against apparent magnitude in B band, (e) the common logarithm of H{\sc{i}} mass with a bin of 0.1 dex, (f) H{\sc{i}} line velocity width with a bin of 10 $km s^{-1}$, (g)H{\sc{i}} mass against B-band luminosity (measured by our own photometry), and (h) central surface brightness against H{\sc{i}} mass \label{fig:fig11}.}
\end{figure*}

We derived the radio information of all the galaxies in our LSB galaxy sample from the ALFALFA catalogue and showed the distributions of them in Figure ~\ref{fig:fig11} (e)$\sim$(h).

As shown in Figure ~\ref{fig:fig11} (e), the common logarithm of H{\sc{i}}-gas mass (log $M(H{\sc{i}})/M\odot$) of the LSB galaxy sample ranges from 6.11 to 10.36 dex, with 32$\%$ (365/1129) having high mass of H{\sc{i}} gas (log M(H{\sc{i}})/M$\odot\geq$ 9.5 dex), 63$\%$ (712/1129) having medium mass of H{\sc{i}} gas (log M(H{\sc{i}})/M$\odot$- 7.7$\sim$9.5 dex), and the left 5$\%$ (52/1129) galaxies having low mass of H{\sc{i}} gas (log M(H{\sc{i}})/M$\odot \leq$ 7.7 dex; \citet{Huang12}). Such a distribition indicates that LSB galaxies are mostly rich in gas with high or medium mass of H{\sc{i}}. The velocity width of H{\sc{i}} line (W$_{50}$) covers from 11.0 to 443.0 $km s^{-1}$ (Figure ~\ref{fig:fig11} (f)), from which only 40$\%$ (450/1129) have narrow H{\sc{i}} line velocity width (W$_{50} <$ 80 $km s^{-1}$; \citet{Huang12}). The distribution of ratios of H{\sc{i}} mass and B-band luminosity (measured by our own photometry) are depicted in Figure ~\ref{fig:fig11} (g). Our LSB galaxy sample has a median M(H{\sc{i}})/L$_{B}$ = 0.87, which suggests that most LSB galaxies of this sample are gas-rich. In Figure ~\ref{fig:fig11} (h), we show the distributions of the B-band central surface brightness against H{\sc{i}} mass.

\subsubsection{Environmental properties}
Evironment is a critical factor which affects the star formation and evolution of galaxies. In this subsection, we made a simple study on the environment where galaxies in our LSB galaxy sample live.

\begin{figure*}
 \centering
 \includegraphics[scale=0.55]{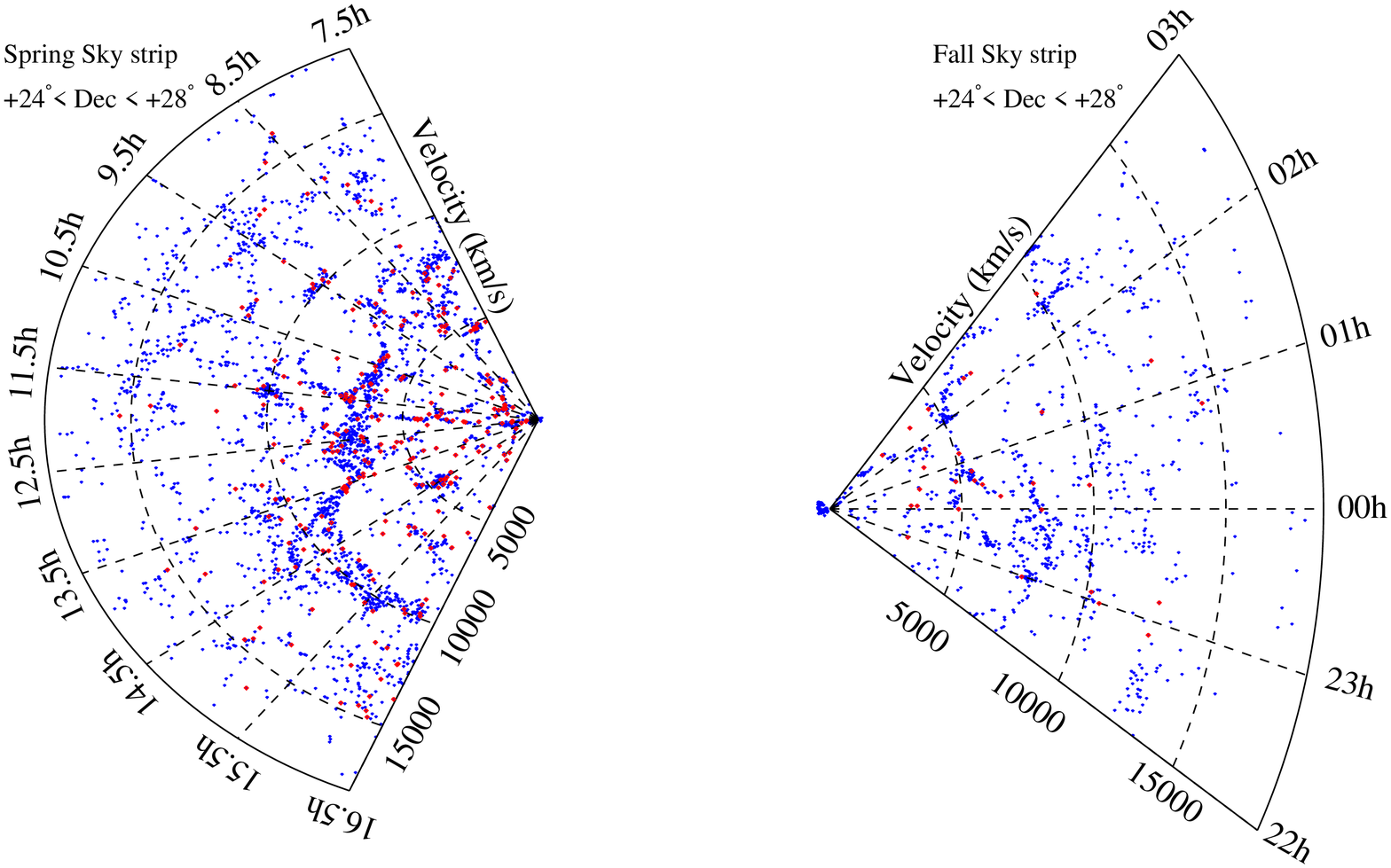}
 \caption[]{ Cone diagrams showing the distribution of $\alpha$.40 H{\sc{i}} sources (blue open circles) and the galaxies of our LSB galaxy sample (red open circles) within both the spring and fall sky strip covering +24$^{\circ} <$decl.$<$+28$^{\circ}$. The diagram shows the volume extending over the full ALFALFA velocity range to 18000 km$\ $s$^{-1}$. \label{fig:fig12}.}
\end{figure*}

In order to have a overview, we provide graphic illustrations of distributions of our LSB galaxies overplotted on the large-scale structure plot (Figures 5 and 6 in \citet{Haynes11}) of the ALFALFA survey in the local universe. As an example, Figure ~\ref{fig:fig12} in our paper shows cone diagrams of a four-degree wide slice of the LSB and $\alpha$.40 galaxies in the ALFALFA spring and fall sky centered on decl.= +26$^{\circ}$, including the full ALFALFA bandpass redshift range cz $<$ 18000 km$\ $s$^{-1}$. Blue open circles mark the locations of galaxies detected by ALFALFA survey, while red open circles denote galaxies in our LSB galaxy sample.

\begin{figure*}
 \centering
 \includegraphics[scale=0.65]{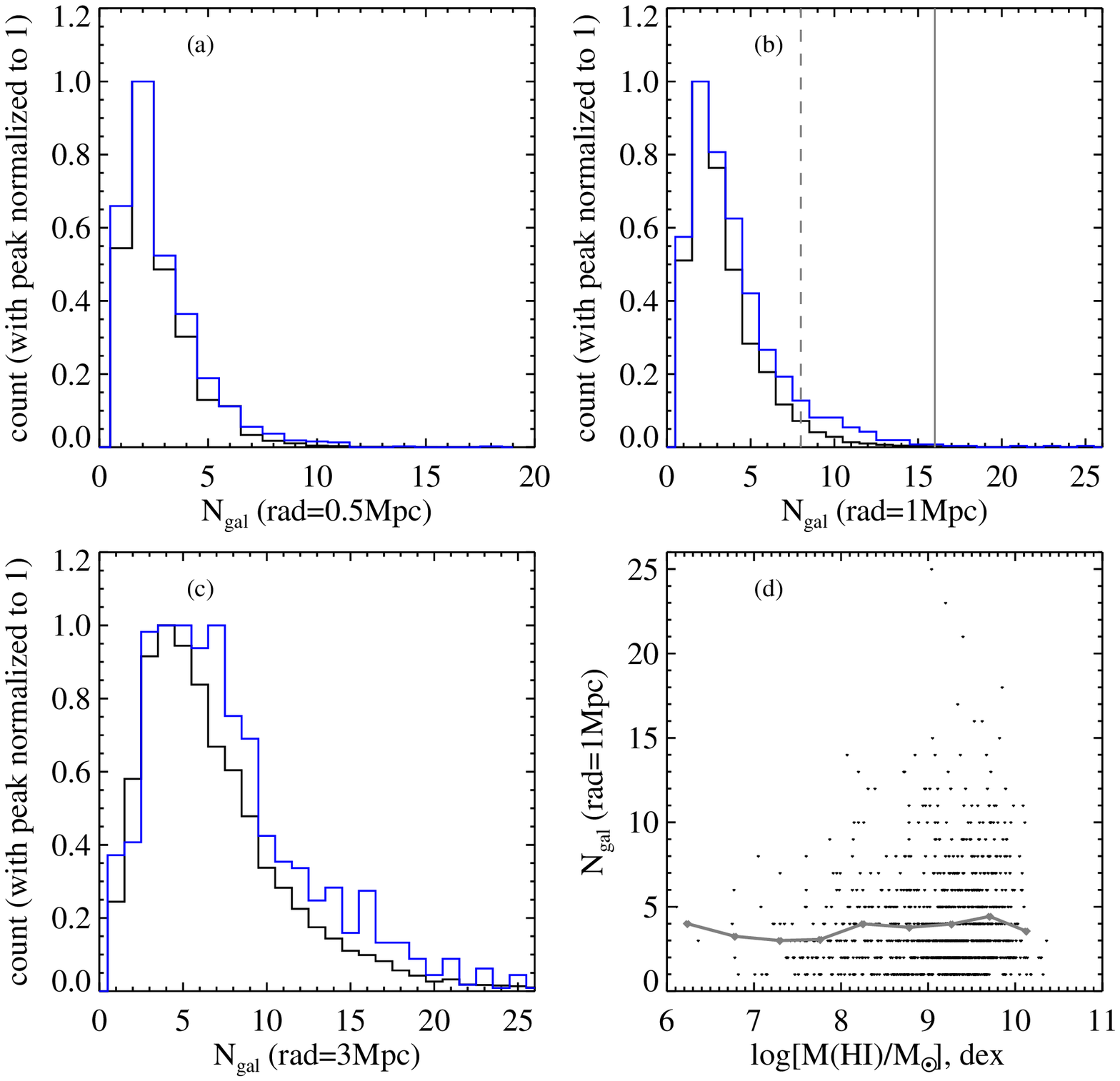}
 \caption[]{Distributions of detectable galaxy number counts by SDSS DR7 within a radius of 0.5 Mpc (a), 1.0 Mpc (b), and 3.0 Mpc (c) and a redshift gap of z $\pm$ 0.04(1+z) for our LSB galaxy sample (blue) comparing with all the $\alpha$.40 galaxies (black). We normalize both peaks of blue and black to 1.0 so as to make them easily compared. The grey dashed is representing N$_{gal}$=8 which is the lower limit of the neighbouring galaxy counts of a galaxy candidate in a cluster, and the grey solid line is representing N$_{gal}$=16 which is the peak of the neighbouring galaxy count distribution for galaxy candidates in clusters. We also show distribution of the detectable galaxy number count with a radius of 1.0 Mpc against the H{\sc{i}} mass for this LSB galaxy sample in panel (d), in which the grey broken line represents the trend between mean H{\sc{i}} mass and mean number count of neighbouring galaxies in 9 H{\sc{i}} mass bins with a binsize of 0.5 dex in log M(H{\sc{i}})/M$\odot$. \label{fig:fig13}.}
\end{figure*}

To study the environment of galaxies in our LSB galaxy sample quantitatively, we follow the method of finding clusters in \citet{Wen09}. For each galaxy in our LSB galaxy sample at a given redshift, $z$, we counted the number of all galaxies detected by the photometric observation of SDSS DR7 within a radius of 1.0 Mpc from the center represented by the galaxy postion and a photometric redshift gap between $z \pm 0.04(1+z)$. Fortunately, the SDSS provides two useful functions for users. One is the $fGetNearbyObjAllEq$ function which returns a table of all objects detected by SDSS within a radius in $arcmin$ of a given equatorial point, and the other is the $fCosmoDa$ function which returns the angular diameter distance at a given redshift. So with the help with these functions, we can easily derive the number of all objects within that radius and redshift space detected by SDSS photometric survey. We show the simple environmental properties of our LSB galaxy sample in Figure ~\ref{fig:fig13}, which shows the distributions of the number counts of galaxies within a radius of 0.5 Mpc in panel (a), 1.0 Mpc in panel (b), and 3.0 Mpc in panel (c) and scatter distribution of the detectable galaxy number count with a radius of 1.0 Mpc against the H{\sc{i}} mass for this LSB galaxy sample in panel (d).Obviously in Figure ~\ref{fig:fig13}(b), about 92$\%$ out of the total galaxies in our LSB galaxy sample have less than 8 galaxies around within the 1.0 Mpc radius and $z \pm 0.04(1+z)$ redshift space. It strongly distinguishes from the count distribution of the member galaxy candidates of clusters within 1.0 Mpc in Figure 4 in \citet{Wen09}, which distributes totally from greater than 8 (grey dashed line) to 50 with the peak at 16 (grey solid line), comparing to the peak of 2$\sim$3 for our LSB galaxy sample. Even if the radius is increased to 3.0 Mpc, the portion of galaxies having less than 8 detectable neighbours is still as high as 65$\%$. This definitely indicates that LSB galaxies are more likely to reside in the field environment. Additionally, from Figure ~\ref{fig:fig13}(a),(b) and (c), the parent sample has nearly the same number of neighbours as the LSB sample. This is large because that the parent sample itself is a gas-rich galaxy sample from the ALFALFA H{\sc{i}} survey as galaxies rich in gas are generally favors low-density environment. In Figure ~\ref{fig:fig13}{d}, we show distribution of the neighbouring galaxy counts againt the H{\sc{i}} mass for our LSB galaxy sample as black dots. As mentioned before, the H{\sc{i}} mass of our LSB galaxy sample ranges from 6.5 to 11.0 dex in log M(H{\sc{i}})/M$\odot$. For a clear understanding of the probable trend between the count of neighbouring galaxies and the H{\sc{i}} mass, we divided the H{\sc{i}} mass range of the LSB galaxy sample into 9 bins with a binsize of 0.5 dex in log M(H{\sc{i}})/M$\odot$, and investigated the relation (the grey broken line in Figure ~\ref{fig:fig13}{d}) between the mean H{\sc{i}} masses and mean number counts of neighbouring galaxies of the 9 mass bins, which does not give a visible relation between the mass and mean counts of neighbouring galaxies, but shows that the mean counts of neighbouring galaxies in all massbins appear lower than 5.

\section{\textbf{Discussion: impact of bulges on fitting results}}

\begin{figure*}
 \centering
 \includegraphics[scale=0.6]{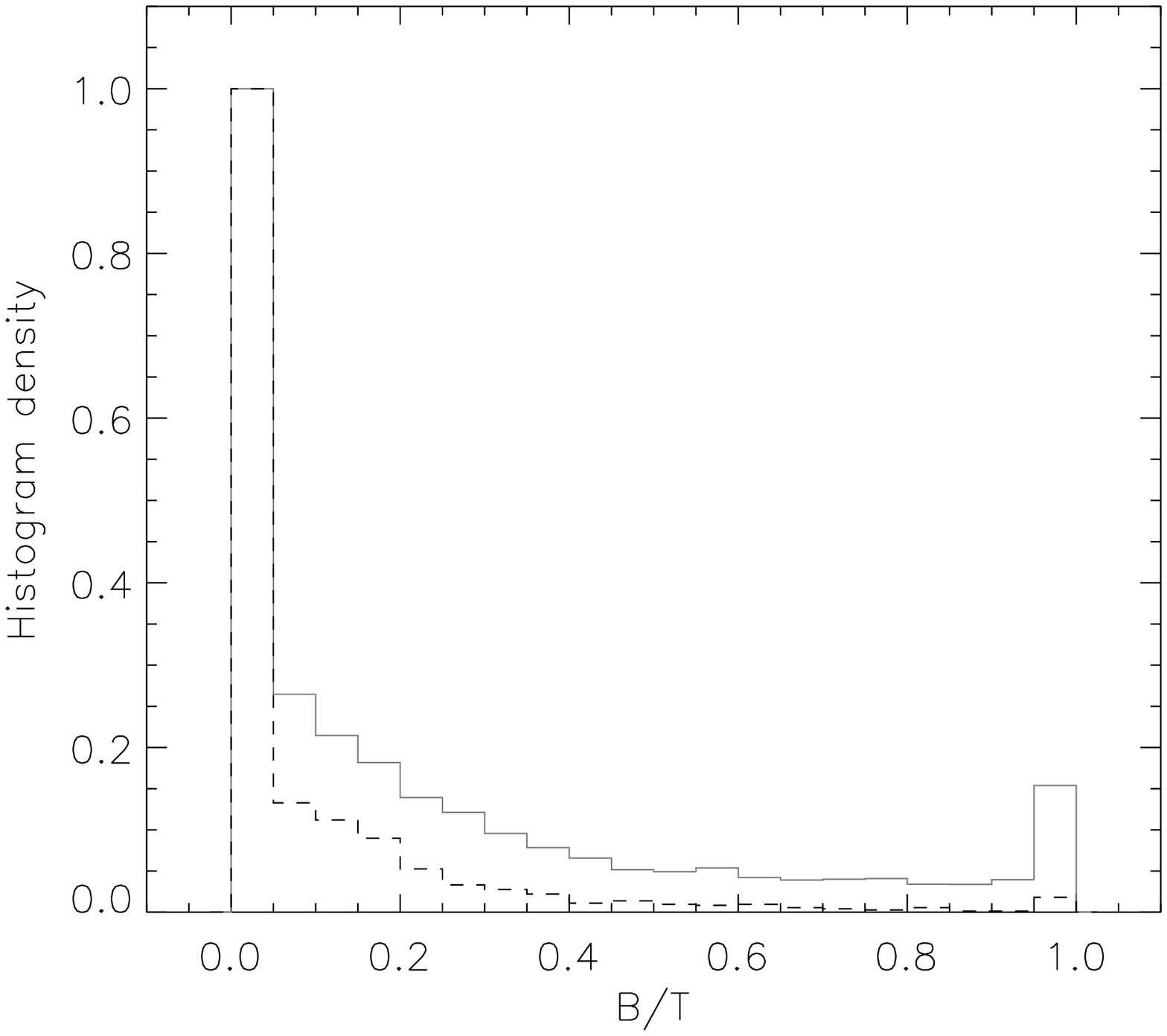}
 \caption[]{Distribution of the bulge-to-total ratio (B/T) of our LSB galaxy sample. Here, the B/T  is derived from the parameter ``fracDev$_{r}$"  from the SDSS DR7 catalogue which represents the fraction of r-band luminosity contributed by the bulge relative to the luminosity contributed by all components (bulge + disk) of galaxies. The dashed black show the bulge-to-total ratios of the 1129 galaxies of our LSB galaxy sample. For a comparison, the B/T distribution of the 12423 galaxies of our parent sample is overplotted as solid grey. Both histograms are normalized to have a maximum value equal to 1.0 for a clear comparison.  \label{fig:fig15}}
\end{figure*}

As LSB galaxies are assumed to be disc-dominated and lack strong bulges, we only adopt a single exponential function as the model in Section 3 to perform a fast search of LSB galaxies from our parent sample by Galfit. This fast method of using only a single exponential profile during the Galfit fitting has already been used  by \citet{Trachternach06} to search LSB galaxies from gas-rich galaxies of a blind optical survey. This fast method tends to be good at detection of the disc-dominated LSB galaxies. However, it may preclude detection of the bulge-dominated LSB galaxies. In Figure ~\ref{fig:fig15}, we show the distribution of the bulge-to-total ratio (B/T) for 1129 galaxies of our LSB galaxy sample (dashed black), and for a comparison the B/T distribution of all of the 12423 galaxies of our parent sample is overplotted (solid grey). Both histograms are normalized to have a maximum value equal to 1.0.  It is worth noting that these B/T values are derived from the parameter "fracDev$_{r}$" from the SDSS DR7 catalogue, which is defined as the ratio of luminosity contributed by the bulge relative to the total luminosity of galaxies in r band. As is once proposed that  B/T = 0.4 can be roughly considered as a division between late-type (disc) and early-type (E/S0) galaxies \citep{Simien86, Li07}, from Figure ~\ref{fig:fig15}, it can be concluded  that among the 1129 galaxies of our LSB galaxy sample,  50.6$\%$ (571 galaxies) are pure disc galaxies (B/T =0), 43.6$\%$ (492 galaxies) are disc-dominated galaxies (0$<$B/T$\leq$0.4) and only 5.8$\%$ (66 galaxies) are bulge-dominated galaxies (0.4$<$B/T$\leq$1.0). Comparing with the composition ratio of our parent sample, 26.3$\%$, 50.2$\%$ and 23.5$\%$ of which are respectively pure disc, disc-dominated and bulge-dominated galaxies, our LSB galaxy sample is more of a disc-dominated LSB galaxy sample that is deficient in the bulge-dominated LSB galaxies.

\begin{figure*}
 \centering
 \includegraphics[scale=0.5]{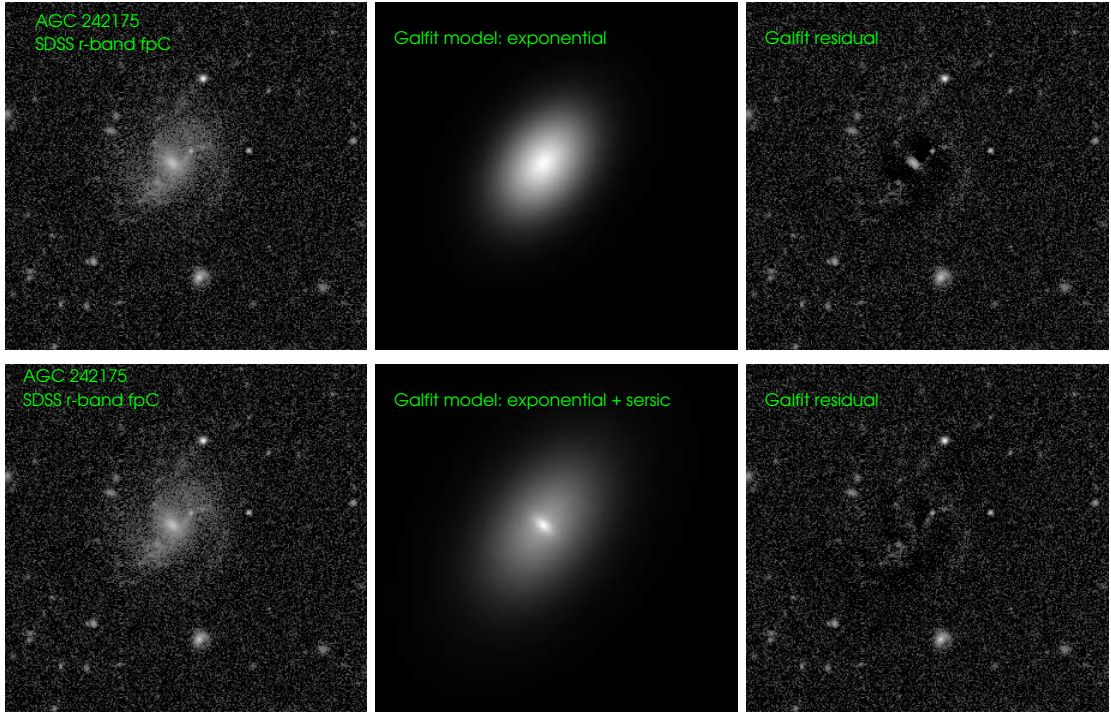}
 \caption[]{Galfit fitting by a single component and a combine of two components. The top panel shows the exponential profile study for the galaxy AGC 242175 in our LSB galaxy sample, and the bottom panel shows the two-component (exponential + S$\grave{e}$rsic) study for it. From left to right are respectively the SDSS r-band image, the final model of the galaxy and the residual image formed by subtracting the final model from the r-band image. \label{fig:fig16}}
\end{figure*}

     In this disc-dominated LSB galaxy sample, it is correct to fit the radial profile of the pure disc galaxies with only a single exponential profile. However, for the galaxies that do have bulges in this LSB galaxy sample, the derived central disk surface brightnesses might be biased towards brighter values by fitting only with a single exponential profile. Instead, a two component (S$\grave{e}$rsic + exponential) study would be much better for these LSB galaxies with bulges. To carry out a test, we randomly take 20 galaxies with different levels of bulges (0.05 $<$ B/T $<$1.0) and morphologically apparent disks from our LSB galaxy sample as the test sample. Then, for this test sample, we fit the galaxies one by one by Galfit with a combine of two components which are the disk component expressed by an exponential radial profile and the bulge component expressed by a S$\grave{e}$rsic radial profile. Comparing with the previous results of fitting these galaxies by Galfit with only a single exponential radial profile, the two component fitting results seem to be better, especially for the galaxies with morphologically larger bulges because the bulges of these galaxies can be subtracted more cleanly as can be seen from the residual images of the two component fitting results. In Figure ~\ref{fig:fig16}, we show images of both the single component fitting (top panel) and two component fitting (bottom panel) by the Galfit for one galaxy with B/T equal to 0.61 from the test sample for an example. As can be obviously seen from Figure ~\ref{fig:fig16}, the two component study can fit the bulge of the galaxy more sufficiently than the fitting only by a single exponential profile. 

\begin{figure*}
 \centering
 \includegraphics[scale=0.5]{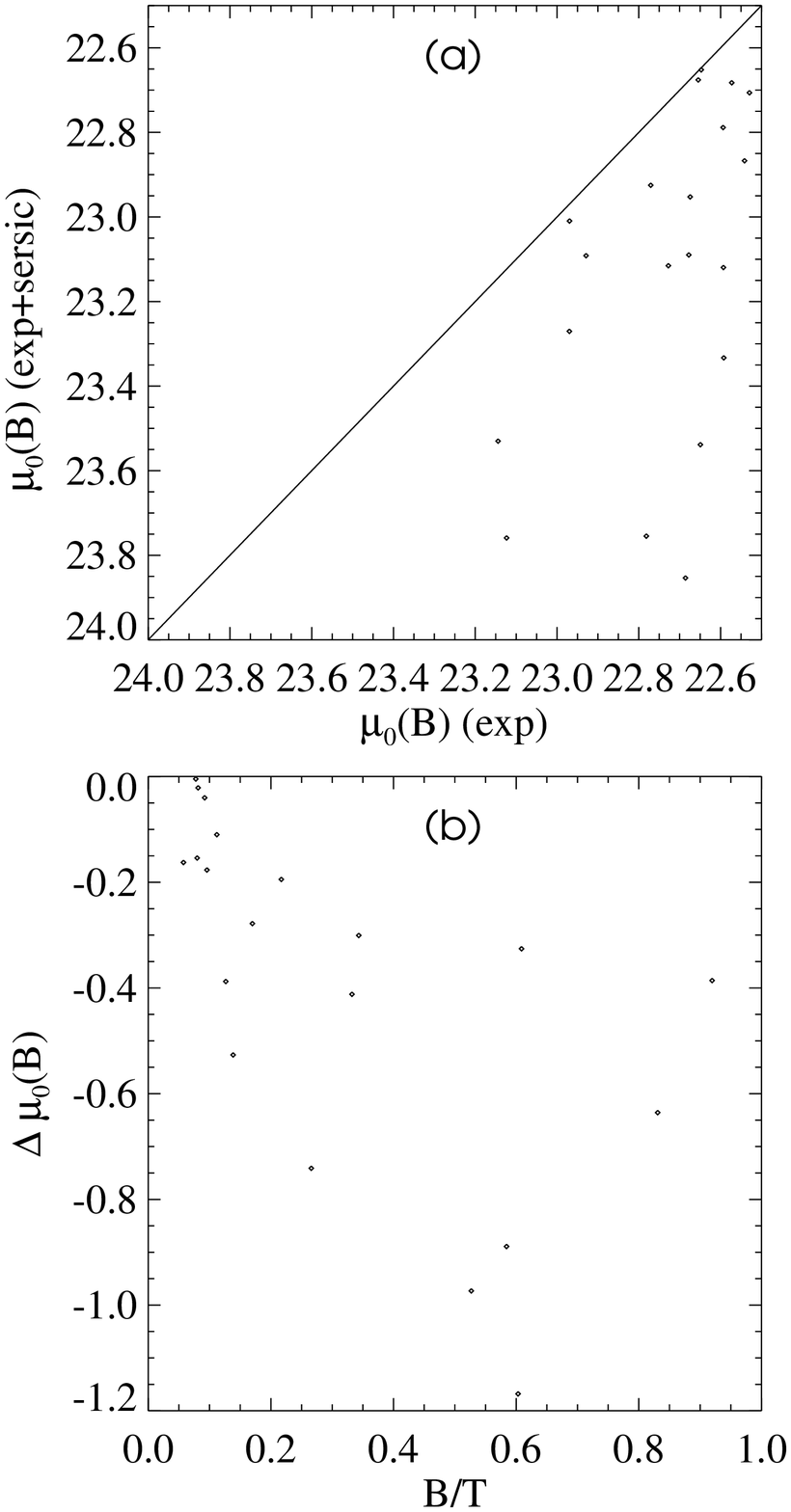}
 \caption[]{Comparison of central disk surface brightnesses of 20 galaxies derived respectively from a single component and a combine of two components fitting. The top panel (a) shows the comparison of the B-band central disk surface brightnesses derived by a single exponential fitting, $\mu_{0}$(B) (exp) with those derived by a two component of exponential and S$\grave{e}$rsic (exp + sersic) profile fitting $\mu_{0}$(B)(exp+sersic). The bottom panel (b) shows the differences of $\mu_{0}$(B)(exp+sersic) from $\mu_{0}$(B) (exp) versus the bulge-to-total luminosity ratio (B/T). The surface brightness is given in units of mag$\ arcsec^{-2}$. \label{fig:fig17}}
\end{figure*}

Being affected by the existing central bulge, the central disk surface brightness would be biased to a brighter value if only the single exponential profile is used to fit this galaxy.  The top panel of Figure ~\ref{fig:fig17} shows comparisons between the derived central disk surface brightnesses for the 20 galaxies of the test sample respectively from the single exponential profile fittting and the two component study by Galfit, which shows that the central disk surface brightnesses are generally overestimated (to brighter values) by fitting only with a single exponential profile. The bottom panel of Figure ~\ref{fig:fig17} shows that the overestimations of central disk surface brightnesses are becoming higher for galaxies with relatively larger bulges.

Our LSB galaxy sample is established basing on the central disk surface brightness derived from the Galfit fitting by only a single exponential profile, so the central disk surface brightnesses of this LSB galaxy sample are overestimated (to brighter values) more or less, especially for those galaxies with large central bulges or light concentrations. \textbf{The stronger the bulges, and the larger the overestimations are (Figure~\ref{fig:fig17}). Nonetheless, galaxies with stronger bulges or light concentrations are not dominant in our LSB galaxy sample (Figure~\ref{fig:fig15}). Quantitatively, from the bottom panel of Figure~\ref{fig:fig17},  for LSB galaxies with weak or no bulges (B/T $<$ 0.2), which contributes 84.7$\%$ (956/1129) to the sample size, the impact of introducing a bulge component on the derived central surface brightness is less than 0.2 mag arcsec$^{-2}$, although the effect can be larger than 0.2 mag arcsec$^{-2}$ for the left 15.3$\%$ fraction (173/1129) of galaxies in our sample when bulges are stronger (0.2$\leq$B/T $\leq$ 1).} 

In the case of overestimation, the true central disk surface brightnesses for those galaxies already in our LSB galaxy sample should be even fainer than currently, so these galaxies currently in our sample should still be members of our LSB galaxy sample. \textbf{Although LSB galaxies are rare to have large bulges or central light concentrations, such bulge-dominated LSB galaxies do exist. However, for such galaxies which should essentially be true LSB galaxies with relatively large bulges or central light concentrations in the parent sample, they tend to be mistakenly identified as non-LSB galaxies only due to the overestimation of the central disk surface brightness (to brighter values  than the LSB galaxy threshold ) by our single exponential profile fitting method instead of a two-component fitting. }  Therefore, we have to acknowledge that our LSB galaxy sample in this paper is more of a disk-dominated LSB galaxy sample and it might lack the true LSB galaxies which have strong bulges or large central light concentrations as a result of the overestimation of central disk surface brightness by using only a single exponential component fitting. \textbf{This can be evidenced by the large contribution (84.7$\%$ ) of disk-dominated galaxies (B/T $<$ 0.2) to our LSB galaxy sample}.  

Although the fitting method of using only a single exponential profile tends to preclude the detection of LSB galaxies which have relatively large bulges or central light concentrations, it is fast and valid for the fast majority of the LSB galaxies \textbf{because LSB galaxies are believed to be mostly disk-dominated galaxies. In our current LSB galaxy sample, the disk-dominated galaxies (B/T $<$ 0.2) contribute up to 84.7$\%$ to the sample size. Additionally, seen from Figure~\ref{fig:fig15}, the B/Ts of this LSB galaxy sample distribute essentially the same way as those of all the parent sample do at the small B/T end. This consistency may also strengthen the validity of this fast method to select the disk-dominated LSB galaxies from all of the disk-dominated galaxies in the parent sample as completely as possible.}

   In view of the above-mentioned fact that our current LSB galaxy sample is a disk-dominated sample and might be deficient in the true bulge-dominated LSB galaxies, in the future, we are planning to make a two-component study for galaxies with bulges or central light concentrations in our parent sample and calculate their central disk surface brightnesses with as possible as small bias. Basing on the new central disk surface brightnesses which are expected to be systematically fainter than the ones derived by only a single exponential profile study, we believe that more true LSB galaxies with bulges or central light concentrations would be picked out and join the LSB galaxy sample. Then, we will sophisticatedly compare our current LSB galaxy sample with the future sample derived by the two-component study, for which it is believed that the contents of Section 4.1 and 4.2 in this paper will change. However, these two sections now could well present the features of our current LSB galaxy sample which is positioned as a non-edge-on disk-dominated LSB galaxy sample selected from the $\alpha$.40-SDSS DR7 survey.  As ALFALFA is a H{\sc{i}} survey, our parent sample should be itself composed of more gas-rich galaxies, so this would surely lead our current LSB galaxy sample towards a sample of having more gas-rich LSB galaxies and lacking sufficient gas-poor LSB galaxies.

\section{Summary}
LSB galaxies are objects with central surface brightnesses at least one magnitude fainter than the night sky.
They are galaxies of a type that encompasses many of the ``extremes" in galaxy properties \citep{O'Neil04} so they play an important role in understanding the galaxy formation and evolution of the uninverse. As most LSB galaxies are rich in gas, the surveys of gas-rich galaxies would be good samples for us to efficiently select LSB galaxies.

  The ALFALFA survey is a blind extragalactic 21cm H{\sc{i}} survey in the local universe (z $\leq$ 0.06) for 7000 deg$^2$ sky area, 40$\%$ of which has already been released to the public as the $\alpha$.40 catelogue.  So the $\alpha$.40-SDSS DR7 sample consisting of 12423 SDSS-primary galaxies provides us one of the best survey combinations to select a relatively unbiased sample of LSB galaxies which are mostly gas-rich in the local universe. Therefore, we expected to define a LSB galaxy sample from the $\alpha$.40-SDSS DR7 sample in this paper.

The selection of LSB galaxies is sensitive to the sky backgrounds. As sky backgrounds are systematically overestimated for galaxies by the SDSS photometric pipeline, especially for those bright galaxies or galaxies with extended low surface brightness outskirts, we accurately re-estimated the sky background of SDSS images in both g and r bands for each galaxy of the $\alpha$.40-SDSS sample, using a careful method of row-by-row and column-by-column fitting. After sky subtraction, we did the surface photometry by the SExtractor software and fitted the geometry by the Galfit software for each galaxy in g and r bands. Basing on both the photometric and geometric results derived by ourselves, the central surface brightness in B band ($\mu_{0}$(B)) could be calculated for every galaxy. Then we selected galaxies with $\mu_{0}$(B) $>$ 22.5 mag arcsec$^{-2}$ and the axis ratio b/a $>$ 0.3 to establish a sample of LSB galaxies from the $\alpha$.40-SDSS DR7 survey.

This LSB galaxy sample, consisting 1129 galaxies from the $\alpha$.40-SDSS DR7 sample, is a relatively unbiased sample of LSB galaxies and it is complete in both the H{\sc{i}} observation and the optical magnitude within the limit of SDSS photometric survey.  This LSB galaxy sample spans from 22.5 to 30.0 for $\mu_{0}$(B) with a large fraction (87$\%$) in 22.5 $< \mu_{0}$(B) $\leq$ 24.0 mag arcsec$^{-2}$, 9$\%$ in 24.0 $< \mu_{0}$(B) $\leq$ 25.0 mag arcsec$^{-2}$, and the left 4$\%$ fainter than 25.0 mag arcsec$^{-2}$, from -27.0 to -12.5 mag for the absolute magnitude in B band (M(B)) with 43 faintest galaxies (M(B) $>$ -17.3 mag), from -0.2 to 1.9 mag for the B-V color with 98.4$\%$ bluer than B-V=0.75 mag indicating that it is a blue LSB sample. In the aspect of H{\sc{i}} properties, a large portion (63$\%$) of the sample has high mass of H{\sc{i}} (M(H{\sc{i}}) $>$ 10$^{9.5}$M$\odot$) and only a few (1$\%$) has very low mass of H{\sc{i}} (M(H{\sc{i}}) $<$ 10$^{7.7}$M$\odot$). These indicate that members of this LSB galaxy sample are mostly gas-rich, with a median M$_{H{\sc{i}}}$/L$_{B} > $1. Additionally, we investigated the environment of galaxies in this LSB galaxy sample. We counted the number of all neighbours of the central galaxy detected by the SDSS photometric survey within a radius of 0.5Mpc, 1.0 Mpc and 3Mpc and a photometric redshift gap between z $\pm$ 0.04(1 + z) and made a statistic on the number counts of neighbours. The distributions of the neighbour counts strongly evidenced that LSB galaxies prefer to reside in the environment of low density, comparing with the neighbouring galaxy counts of cluster candidates shown in \citet{Wen09}.

However, this LSB galaxy sample has its own drawback. As discussed in Section 5, the selection of this LSB galaxy sample is basing on the central disk surface brightnesses, which are derived by performing only a single exponential profile fit to the galaxies of the $\alpha$.40-SDSS sample. This fitting method of only using a single exponential model is fast and valid for the fast majority of the LSB galaxies. However, it indeed overestimates the central disk surface brightnesses for galaxies with bulges or central concentrations of light by comparing with the results of a two-component study for these galaxies in our LSB galaxy sample. \textbf{For LSB galaxies with weak or no bulges (B/T $<$ 0.2), which contributes 84.7$\%$ (956/1129) to the sample size, the impact of introducing a second component (S$\grave{e}$rsic) on the derived central surface brightness is less than 0.2 mag arcsec$^-2$, although the effect can be larger than for the left 15.3$\%$ fraction (173/1129) of galaxies in our sample which have stronger bulges (0.2$\leq$B/T $\leq$ 1). The stronger the bulges, and the larger the effect will be. Nonetheless, fraction of galaxies with very strong bulges (5.8$\%$ with B/T=1)is very small in our LSB galaxy sample.} As the present central disk surface brightnesses systematically bias towards brightner values and should be fainter than currently after a two-component study for galaxies with bulges in our LSB galaxy sample, our LSB galaxy sample is a sample consisting of more disk-dominated galaxies but being deficient in the true bulge-dominated LSB galaxies which are rare but do exist.  In a word, the LSB galaxy sample defined in this paper is a relatively unbiased sample of gas-rich, non-edge, and disk-dominated LSB galaxies from the overlap between the ALFALFA 21cm H{\sc{i}} survey and the SDSS DR7.




\acknowledgments
We would like to thank the referee for the helpful suggestions. We also thank ALFALFA team for providing the $\alpha$.40 catalog and the SDSS team for the wonderful released SDSS fpC images. This project is supported by the National Natural Science Foundation of China (Grant No.11403037), the China Ministry of Science and Technology under the State Key Development Program for Basic Research (2012CB821800, 2014CB845705), the National Natural Science Foundation of China (Grant Nos.11303038, 11225316, 11173030, 11078017, 10810301043, 10773014 and 10273012),  the Strategic Priority Research Program "The Emergence of Cosmological Structures" of the Chinese Academy of Sciences(Grant No. XDB09000000), the Collaborative Innovation Center of Modern Astronomy and Space Exploration and the Key Laboratory of Optical Astronomy, the National Astronomical Observatories, Chinese Academy of Sciences.






\appendix





\clearpage



\clearpage









\clearpage


\clearpage



\clearpage




\end{document}